\documentclass[
  prb,
  superscriptaddress,
  reprint
  ]{revtex4-1}
\usepackage[utf8]{inputenc}
\usepackage{CJKutf8}
\usepackage[T1]{fontenc}
\usepackage[usenames, dvipsnames]{color}
\usepackage{graphicx}
\usepackage{float}
\usepackage{empheq}
\usepackage{textcomp}
\usepackage{bm}
\usepackage{verbatim}
\usepackage{amsmath}
\usepackage{mathrsfs} 
\usepackage{amssymb}
\usepackage{amsthm}
\usepackage{xfrac}
\usepackage{hyperref}
\usepackage{physics}
\usepackage{dsfont}

\graphicspath{ {images/} }

\newcommand*{\Neel}{}
\def\Neel/{N\'eel}
\newcommand*{\Schrodinger}{}
\def\Schrodinger/{Schr\"odinger}

\newcommand*{\bbid}{\ensuremath{\mathds{1}}}

\newcommand*{\bmag}{\ensuremath{\mathbf{m}}}
\newcommand*{\bn}{\ensuremath{\mathbf{n}}}
\newcommand*{\bN}{\ensuremath{\mathbf{N}}}
\newcommand*{\bx}{\ensuremath{\mathbf{x}}}
\newcommand*{\bA}{\ensuremath{\mathbf{A}}}
\newcommand*{\bB}{\ensuremath{\mathbf{B}}}
\newcommand*{\bD}{\ensuremath{\mathbf{D}}}
\newcommand*{\bE}{\ensuremath{\mathbf{E}}}

\newcommand*{\ba}{\ensuremath{\mathbf{a}}}
\newcommand*{\bb}{\ensuremath{\mathbf{b}}}
\newcommand*{\ma}{\ensuremath{\mathbf{m}_A}}
\newcommand*{\mb}{\ensuremath{\mathbf{m}_B}}
\newcommand*{\fraka}{\ensuremath{\hat{a}}}
\newcommand*{\balpha}{\ensuremath{\bm{\alpha}}}
\newcommand*{\bbeta}{\ensuremath{\bm{\beta}}}
\newcommand*{\bEta}{\ensuremath{\bm{\eta}}}

\newcommand*{\bk}{\ensuremath{\mathbf{k}}}
\newcommand*{\bq}{\ensuremath{\mathbf{q}}}
\newcommand*{\dbk}{\ensuremath{\dd \mathbf{k}\,}}
\newcommand*{\dbq}{\ensuremath{\dd \mathbf{q}\,}}

\newcommand*{\bqc}{\ensuremath{\mathbf{q}_c}}
\newcommand*{\bkc}{\ensuremath{\mathbf{k}_c}}
\newcommand*{\bxc}{\ensuremath{\mathbf{x}_c}}

\newcommand*{\bOmega}{\ensuremath{\bm{\Omega}}}

\newcommand*{\tma}{\ensuremath{\tilde{\mathbf{m}}_A}}
\newcommand*{\tmb}{\ensuremath{\tilde{\mathbf{m}}_B}}
\newcommand*{\calA}{\ensuremath{\mathcal{A}}}
\newcommand*{\calG}{\ensuremath{\mathcal{G}}}

\newcommand*{\calZ}{\ensuremath{Z}}
\newcommand*{\calJ}{\ensuremath{J}}
\newcommand*{\covA}[1]{\ensuremath{\mathbb{A}_{#1}}}

\newcommand*{\calJaf}{\ensuremath{\calJ _\text{AF}}}
\newcommand*{\calJf}{\ensuremath{\calJ _\text{F}}}

\definecolor{basebg}{RGB}{0, 43, 54}
\definecolor{basefg}{RGB}{131, 148, 150}
\definecolor{colormmablue}{RGB}{94, 129, 181}
\definecolor{colormmaorange}{RGB}{225, 156, 36}
\definecolor{colormmagreen}{RGB}{143, 176, 50}
\begin{document}

\title{Nonabelian magnonics in antiferromagnets}

\begin{CJK*}{UTF8}{gbsn}
\author{Matthew W.~Daniels} 
  \email[Corresponding author:~]{danielsmw@protonmail.com}
  \affiliation{ Department of Physics,
                Carnegie Mellon University,
                Pittsburgh, PA 15213, USA }

\author{Ran Cheng}
  \affiliation{ Department of Physics,
                Carnegie Mellon University,
                Pittsburgh, PA 15213, USA }
  \affiliation{ Department of Electrical and Computer Engineering,
                Carnegie Mellon University,
                Pittsburgh, PA 15213, USA }

\author{Weichao Yu (余伟超)}
\affiliation{ Department of Physics and State Key
              Laboratory of Surface Physics,
              Fudan University,
              Shanghai 200433, China}

\author{Jiang Xiao (萧江)}
\affiliation{ Department of Physics and State Key
              Laboratory of Surface Physics,
              Fudan University,
              Shanghai 200433, China}
\affiliation{ Collaborative Innovation Center
              of Advanced Microstructures,
              Nanjing 210093,
              China}
\affiliation{ Institute for Nanoelectronics Devices and Quantum Computing, Fudan University, Shanghai 200433, China }

\author{Di Xiao}
  \affiliation{ Department of Physics,
                Carnegie Mellon University,
                Pittsburgh, PA 15213, USA }

\begin{abstract}
  We present a semiclassical formalism for antiferromagnetic (AFM) magnonics
  which promotes the central ingredient of \emph{spin wave chirality}, encoded
  in a quantity called magnonic isospin, to a first-class citizen of the theory.
  We use this formalism to unify results of interest from the field under a
  single chirality-centric formulation. Our main result is that the isospin is
  governed by unitary time evolution, through a Hamiltonian projected down from
  the full spin wave dynamics. Because isospin is SU(2)-valued, its dynamics on
  the Bloch sphere are precisely rotations---which, in general, do not commute.
  Consequently, the induced group of operations on AFM spin waves is nonabelian.
  This is a paradigmatic departure from ferromagnetic magnonics, which operates
  purely within the abelian group generated by spin wave phase and amplitude.
  Our investigation of this nonabelian magnonics in AFM insulators focuses on
  studying several simple gate operations, and offering in
    broad strokes a program of study for interesting new logic families in
    antiferromagnetic spin wave systems.
\end{abstract}

\maketitle
\end{CJK*}


\section{Introduction}

\begin{figure}
  \centering \includegraphics[width=\columnwidth]{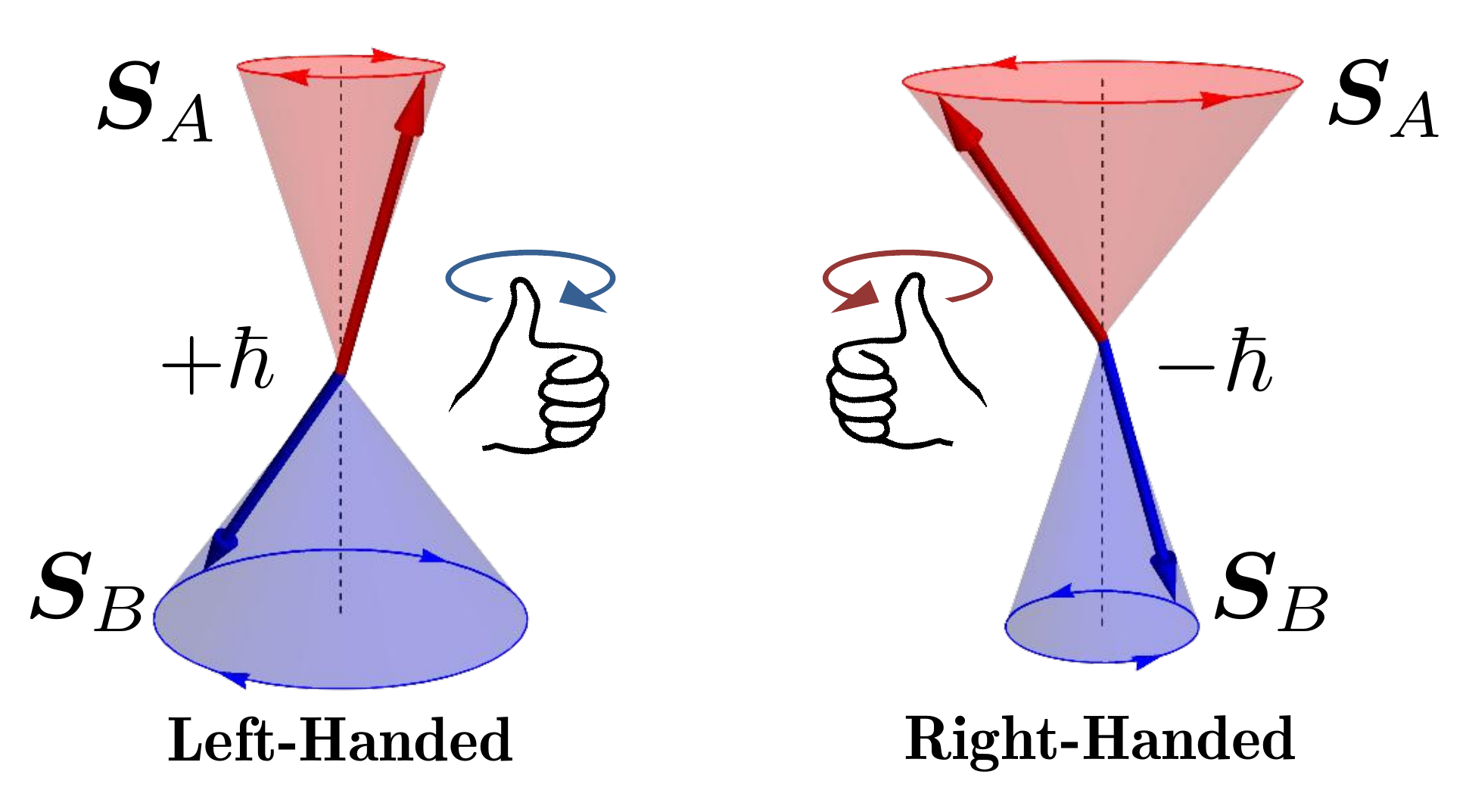}
  \caption{Schematic representations of right-handed and left-handed modes. Red
    and blue arrows demonstrate the spin precession on each of the two
    sublattices. Because the $\hat S_z$ components differ between the
    sublattices during a spin wave precession, each eigenmode carries an
    opposite sign of spin angular momentum.}
  \label{fig:lr}
\end{figure}

Recent years have seen a surge of interest in the generation and in-flight
manipulation of magnons in antiferromagnets (AFMs). We now know that AFM magnons
can couple to the angular momentum carried by
electrons,~\cite{Cheng:2014aa,Daniels:2015aa} photons,~\cite{kanda:2011aa,
  kimel:2005aa, satoh:2010aa} and other spin carriers. Detection of
magnon-mediated spin signals from AFM insulators, typically measured through the
inverse spin Hall effect, has also matured to the point of experimental
implementation.~\cite{Tokura:2015aa, shiomi:2017ab, shiomi:2017aa} It has been
shown that AFM spin waves possess pointed dynamical distinctions from their
ferromagnetic (FM)
counterpart,~\cite{Gitgeatpong:2017tc,cheng:2017qm,Cheng:2016kv} especially in
the presence of spin
texture~\cite{lan:2017aa,Tveten:2014aa,Kim:2014aa,Seki:2015es,Kim:2015el,qaiumzadeh:chiral:2018}
or broken inversion symmetry.~\cite{Cheng:2016kv, Cheng:2016ku, zyuzin:2016aa,
  lan:2017aa} In particular, collinear AFMs possess two degenerate spin wave
eigenmodes of opposite chirality.~\cite{Keffer:1952jw} They are often referred
to as right- and left-handed modes, according to the precessional handedness of
the N\'eel vector (Fig.~\ref{fig:lr}). This notion of \emph{spin wave chirality}
has proved to be a useful narrative tool for understanding how AFM magnonics
differs from the ferromagnetic (FM) case.

As a patchwork of novel results begins to populate the field of AFM magnonics, a
coherent framework for understanding their similarities, differences, and
possible extensions becomes necessary. Our central thesis is that many of these
results can be understood in terms of spin wave chirality, through a spinor
(SU(2)-valued) quantity we refer to as the \emph{magnon isospin}. One important
corollary of this formulation is that---because isospin dynamics proceeds by
intrinsically noncommutative unitary rotations on the Bloch
sphere---implementations of magnonic computing in AFMs will in general be
nonabelian. This fundamental departure from the behavior of FM magnonics calls
for a serious re-investigation of primitive magnonic operations for AFMs;
working only off analogies to extant ferromagnetic proposals is a program
restricted by commutativity, and inevitably lifts only into a small subset of
available AFM computing schemes.

One practical disadvantage of FM magnonics has been the need to constantly
refresh the signal power in a device. This is particularly problematic in
interferometric~\cite{Liu:2011iha, Liu:2011ih, Khitun:2010dx} spin wave logic,
where the boolean output of FM magnonic logic gates is encoded by setting a
threshold amplitude for the spin wave power. Phase interference techniques are
then used to achieve the desired magnon amplitude. Since half of the desired
outputs are represented by suppressing the power spectrum of the magnon signal,
this scheme incurs significant energy inefficiencies and requires sources of
power to constantly refresh the signal.\footnote{Others have proposed using the
  phase to encode information directly,~\cite{Khitun:2010dx} though these
  schemes sometimes end up using spin wave amplifiers anyway. In any case, a
  natural design pattern would seem to use either the phase for computing and
  the amplitude for memory---or vice versa---and so the amplitude will somewhere
  need to be restored from some suppressed state. } Isospin computing resolves
this problem neatly, since we can encode and manipulate data in the spin wave
chirality rather than the spin wave amplitude. This improvement is reminiscent
of proposals for polarization-based optical computing schemes from the
1980s.~\cite{sawchuk:1984aa} A chief practical distinction between AFM isospin
computing and optical computing is that the former can be carried out in
nanoscale solid state systems.

Given the importance of the isospin in AFM magnonics, we consider in this paper
its dynamics for a broad class of interactions that may manifest in AFMs, and
offer an extensible formalism by which others can easily incorporate the effects
of new physical interactions. In the development of this formalism, we find that
there are notable differences between bipartite and synthetic AFMs, and we
discuss the advantages and disadvantages of pursuing non-abelian magnonics in
these two types of systems. We then apply this formalism to a number of
examples, for the threefold purposes of illustrating its use, validating it
against a set of known results, and generating novel results in a few
interesting systems.

With several concrete results in hand, we then propose in broad strokes a
program of next-generation computing based on nonabelian magnonics.\footnote{As
  the magnonic isospin lives in a noncommutative space, device components in an
  isospin computing architecture will not commute in general. \emph{Nonabelian}
  is a synonym for \emph{noncommutative}: hence, the name.} Though FM magnonics
has been studied extensively,~\cite{Khitun:2010dx, Liu:2011ih, Schneider:2008fu,
  Chumak:2015fa} we show that the comparative richness of the AFM isospin offers
dramatically more and different avenues for progress. The fact that isospin
manipulations do not commute offers, by purely algebraic considerations, a more
bountiful landscape for composition of logical operations than can be found in
FMs.

We emphasize that the dual-sublattice nature of AFMs does not \emph{merely}
amount to two copies of FM magnon systems. Though one may be able to import FM
magnonic schemes into the AFM architecture, one could also look to more spinful
classes of physics for inspiration in application. Spintronic
~\cite{Cheng:2016kv} and optical~\cite{lan:2017aa} analogies to AFM magnonics
have proved inspiring for novel device designs. We close by offering
possibilities for future research in this direction.


\section{Formalism}

In this section, we review the AFM spin wave theory in the sublattice formalism,
as we expect many of our readers are more familiar with the
staggered-order-centric approach. We begin by exploring spin wave chirality in a
minimally model: a collinear AFM with easy axis anisotropy. The
  description of easy-axis AFMs such as MnF$_2$, FeF$_2$, or Cr$_2$O$_3$ may
  follow from such a model. Using this familiar context, we review chirality
and the way in which it encodes spin carried by the magnon excitation. We then
review a common formalism for handling spin texture and introduce the
texture-induced gauge fields. Finally, we derive the spin wave equations of
motion in the sublattice formalism by the variational principle. These
subsections set the stage for our main results, which are presented in the next
section.

\subsection{Sublattice-centric magnonics}
\label{sec:two-level-magnonics}
In terms of the two sublattices, the free energy of an easy-axis collinear AFM
in the continuum limit is
\begin{subequations} \label{eq:all}
  \begin{align}
    F &= F_\text{exch} + F_\text{EAA}\label{eq:free-energy-simple}\\
    F_\text{exch} &= \frac{1}{2}\int \calZ \ma\cdot\mb - \calJ \nabla\ma\cdot\nabla\mb \dd[d]{\bx}\label{eq:free-energy-exch}\\
    F_\text{EAA} &= -\frac{K }{2} \int (\ma \cdot \hat z)^2 + (\mb\cdot\hat z)^2\dd[d]{\bx}
  \end{align}
\end{subequations}
Here, $K $ is the easy-axis anisotropy, while $\calZ$ and $\calJ$ are the
so-called homogeneous and inhomogeneous exchange interactions,
respectively.\cite{Tveten:2015aa} They have been chosen so that, under the
change of variables
\begin{equation}
  \bmag = \frac{\ma + \mb}{2} \quad\text{and}\quad \bn = \frac{\ma - \mb}{2},
\end{equation}
the exchange free energy density becomes\footnote{The literature is typically
  consistent with maintaining factors of either $1/2$ or $1$ in front of the
  various terms of Eq.~\eqref{eq:free-energy-m-n}, but avoids mixing them; see
  for instance Refs.~\onlinecite{Tveten:2015aa} or \onlinecite{Hals:2011aa}.
  Since we have expressed the free energy in terms of the sublattices---rather
  than the staggered order $\bn$ and the local magnetization $\bmag$---we will
  instead opt to disperse our factors of $1/2$ symmetrically in
  Eq.~\eqref{eq:free-energy-exch} and require the reader to perform the
  conversion between conventions if needed.}
\begin{equation}
  \mathcal F_\text{exch} = \calZ |\bmag|^2 + \frac{\calJ}{2} |\nabla\bn|^2 + O(|\bmag|^4).\label{eq:free-energy-m-n}
\end{equation}
The quantities $\bmag$ and $\bn$ are the local magnetization and the staggered
order.\footnote{$\bn$ is the order parameter for antiferromagnets, and so AFM
  dynamics are often (though not in the present article) described using $\bn$
  as the primary dynamical variable. Our description of spin wave fields in
  terms of the sublattices, rather than $\bn$, will preserve the first-order (in
  time) nature of the Landau-Lifshitz equation instead of passing over to the
  second-order Klein-Gordon equation governing staggered order fluctuations.
  Maintaining a first-order theory facilitates the use of wavepacket theory that
  we introduce in Sec.~\ref{sec:nonab-wavep-theory}.} We have written
in Eqs.~\eqref{eq:all} a free energy for the classic $g$-type antiferromagnet,
but merely as a convenient concretization. Our main result generalizes to any
kind of collinear AFM order, and in particular we use results for synthetic
AFMs later in the article.

On each sublattice of the AFM, the semiclassical spin dynamics are governed by
the Landau-Lifshitz equation,
\begin{subequations} \label{eq:llgs}
  \begin{align}
    \dot{\bmag}_A &= \ma \times \frac{1}{S}\frac{\delta F}{\delta \ma} \;, \\
    \dot{\bmag}_B &= \mb \times \frac{1}{S}\frac{\delta F}{\delta \mb} \;,
  \end{align}
\end{subequations}
where $F$ is the free energy functional and $S = s\hbar$ the spin magnitude on a
lattice site. Define $\hat z$ as the easy-axis direction, and take the \Neel/
ground state as $\ma = \hat z$ and $\mb = -\hat z$. Spin wave fluctuations, at
linear order in the cone angle by which precessing spins cant away from the
ground state, reside entirely in the $xy$-plane.\footnote{Note that this
  separation into slow and fast variables is exact within linear spin wave
  theory; likewise when we substitute the slow modes $\pm\hat z$ for $\pm R\hat
  z$ come the introduction of spin texture.} It is convenient to write
fluctuations from equilibrium as $\alpha =( m_A^x + i m_A^y)/\sqrt 2$ and $\beta
= (m_B^x + i m_B^y)/\sqrt 2$. Then, in the $\{\alpha,\beta,\alpha^*,\beta^*\}$
basis, the spin wave equations of motion for $\Psi$ are\footnote{Note that
  $\Psi$ is a vector of coefficients for the superposition of fields $\alpha,
  \beta,\alpha^*,\beta^*$---it does not encode the evaluation of the fields
  themselves. It is in this sense that these four variables are free and not
  linked by complex conjugation the way their basis vectors are.}
\begin{equation}
  i (\tau_z\otimes\sigma_z)\dot\Psi = \left(\begin{matrix} \hat h & 0 \\ 0 & \hat h^*\end{matrix}\right)\Psi = \mathcal H \Psi \;,
  \label{eq:simple-4x4}
\end{equation}
where $\tau_j$ are the Pauli matrices in isospin space, $\sigma_j$ the Pauli
matrices in the sublattice subspace, and $\hat h$ is a $2\times2$ hermitian
operator given by
\begin{equation}
  \hat h = \frac{1}{2}\left[ (\calZ + 2K) \bbid_2 + \sigma_x (\calZ + \calJ\nabla^2)\right] \;.
  \label{eq:2-level-ham}
\end{equation}
For the simple free energy we have adopted in Eq.~\eqref{eq:all},
Eq.~\eqref{eq:simple-4x4} apparently contains two copies of the same two-level
dynamics. These two copies are related by complex conjugation, which we write as
the time-reversal operator $\mathcal T$. The mapping of the LLG
  equation onto a \Schrodinger/ equation is standard practice in theoretical
  magnonics,\cite{Dugaev:2005aa,Guslienko:2010kq} but note that our
Eq.~\eqref{eq:simple-4x4} differs from the usual \Schrodinger/ equation by the
appearance of $\tau_z\otimes\sigma_z$ on the left hand side. The mathematical
and philosophical details of \Schrodinger/ equations with this structure have
been considered at length in Ref.~\onlinecite{Mostafazadeh:2003hq}.

Since the Hamiltonian~\eqref{eq:simple-4x4} is block diagonal, let us first
focus on the subspace spanned by $\{\alpha, \beta\}$. Assuming our system is
stationary and translationally invariant, we can make the ansatz $\psi = \psi_0
e^{i(\bk\cdot\bx-\omega t)}$. The resulting eigenproblem is
\begin{equation} \label{eq:hhh} \hbar\omega\sigma_z\psi = \hat h\psi \;.
\end{equation}
For a generic $2\times2$ hermitian operator $\hat h = a\bbid_2 + b\sigma_x +
c\sigma_y + d \sigma_z$, Eq.~\eqref{eq:hhh} has the solution~\cite{Cheng:2016ku}
\begin{equation}
  \psi_0 = \left(\begin{matrix}
      \cosh\frac{\vartheta}{2} \\ -e^{i\varphi} \sinh\frac{\vartheta}{2}
    \end{matrix}\right)
  \quad\text{and}\quad
  \psi_1 = \left(\begin{matrix}
      -\sinh\frac{\vartheta}{2} \\ e^{i\varphi} \cosh\frac{\vartheta}{2}
    \end{matrix}\right) \;,
  \label{eq:two-level-hyperbolic-sols}
\end{equation}
where the angles $\vartheta$ and $\varphi$ are given through
\begin{subequations}
  \label{eq:abc-ltp}
  \begin{align}
    a &= \ell \cosh\vartheta \;, \\
    b &= \ell \sinh\vartheta \cos \varphi \;, \\
    \text{and}\;c &= \ell \sinh\vartheta \sin \varphi \;.
  \end{align}
\end{subequations}
The corresponding eigenvalues are
\begin{equation}
  \hbar\omega = \pm(d + \ell) = \pm \frac{1}{S}\sqrt{\frac{1}{2}\calJ \calZ  k^2 + \calZ  K }
\end{equation}
at leading order in $K$. The well-known resonant energy is given then by
$\hbar\omega_0 = \sqrt{\calZ K }/S$.

We note that the bosonic normalization condition~\cite{colpa:1978aa,
  shindou:2013aa, Cheng:2016ku} $a^2 - b^2 - c^2 = \pm 1$ implies that the space
of Hamiltonians---as well as the eigenvectors themselves---live on the
hyperboloid of two sheets SU(1,1). When $d=0$, as in Eq.~\eqref{eq:2-level-ham},
the eigenvectors have particle-hole symmetry. $\psi_j$ exhibits eigenfrequency
$(-)^j|\omega|$. Analysis of the basis functions shows that $\psi_0$ is a
right-handed precession of $\ma$ (and therefore $\bn$) while $\psi_1$ is a
left-handed precession. We say that they have opposite chirality, namely
right-handed and left-handed chirality.

\begin{figure}
  \centering \includegraphics[width=\columnwidth]{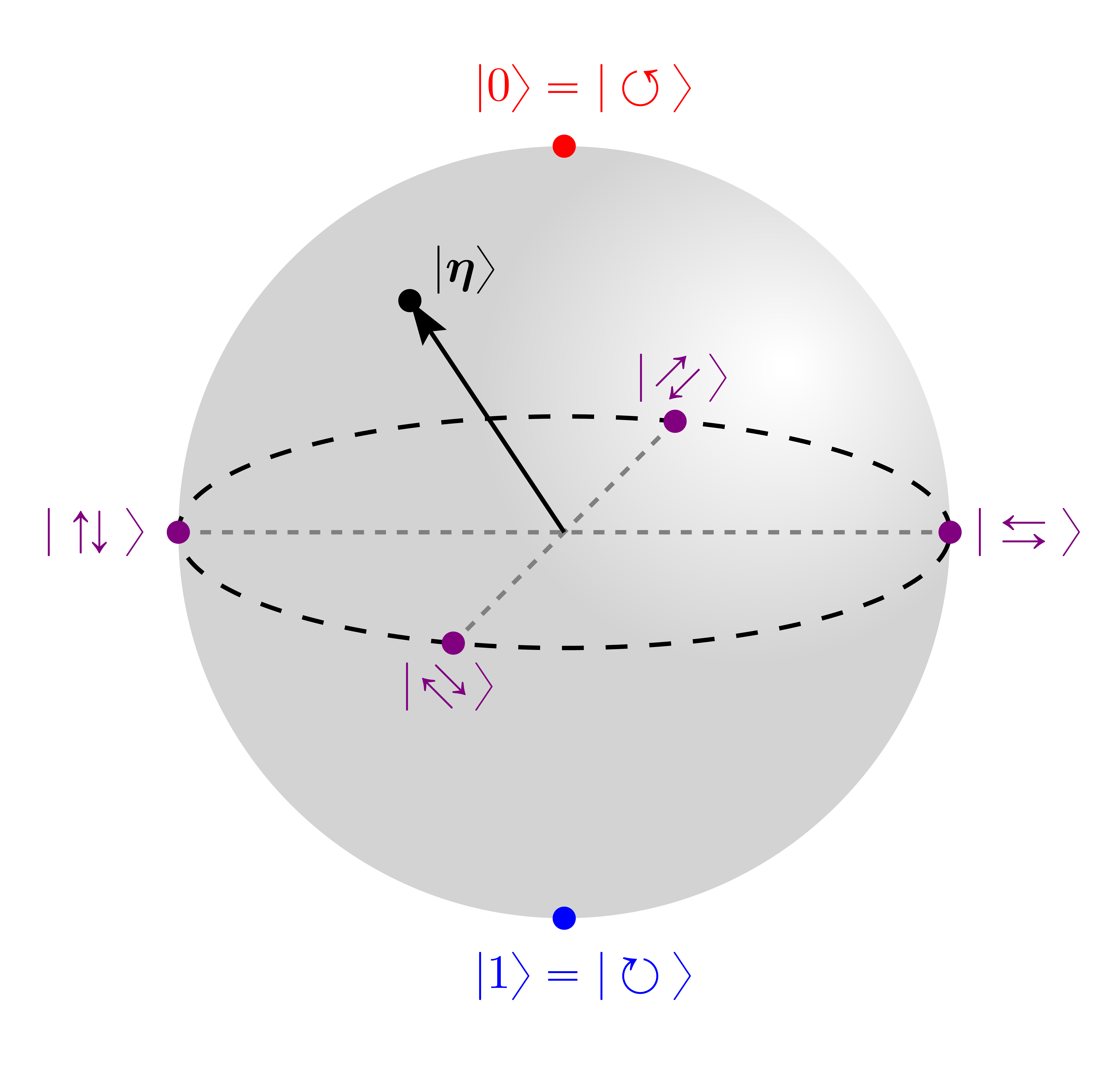}
  \caption{Linear combinations of the right- and left-handed modes $\ket 0 \cong
    \Psi_0$ and $\ket 1 \cong \Psi_1$, respectively, produce an entire Bloch
    sphere's worth of possible isospin states. We have labeled selected states
    by the polarization of the \Neel/ order fluctuations in that state. Right-
    and left-handed modes correspond to right- and left-handed precession of
    $\bn$, while equal linear combinations produce linearly polarized waves. The
    angle of linear polarization depends on the relative phase of the spin waves
    between the sites. Note that $X$ and $Y$-polarized states are orthogonal
    here, while in a traditional quantum spin space $\ket X$ is orthogonal to
    $\ket{-X}$, not $\ket Y$. Since our formalism parameterizes this space in
    terms of a 2-level spinor, we refer to it as a Bloch sphere. Students of
    optics, however, will recognize that it is analogous to the Poincar\'e
    sphere that parameterizes optical polarization states.}
  \label{fig:bloch-sphere}
\end{figure}

Notice that the sister eigenproblem (for $\{\alpha^*, \beta^*\}$) in the lower
two rows of Eq.~\eqref{eq:simple-4x4} has positive frequency solutions
corresponding to left-handed modes and negative frequency solutions
corresponding to right-handed modes. This inversion from the $\{\alpha, \beta\}$
problem arises precisely due to the conjugate basis. We will take the
positive-energy solution from each block,
\begin{equation}
  \Psi_0 = \left(\begin{matrix}
      \cosh\frac{\vartheta}{2} \\ -e^{i\varphi} \sinh\frac{\vartheta}{2} \\ 0 \\ 0
    \end{matrix}\right)
  \quad\text{and}\quad
  \Psi_1 = \left(\begin{matrix}
      0 \\ 0 \\ -\sinh\frac{\vartheta}{2} \\ e^{i\varphi} \cosh\frac{\vartheta}{2}
    \end{matrix}\right),
  \label{eq:four-level-hyperbolic-sols}
\end{equation}
as a chirally-complete basis for the positive energy, degenerate Hilbert
subspace of Eq.~\eqref{eq:simple-4x4}. Note that whereas the solutions
\eqref{eq:two-level-hyperbolic-sols} obey $\mel{\psi_i}{\sigma_z}{\psi_j} =
(-)^j\delta_{ij}$, the solutions $\mel{\Psi_i}{\tau_z\otimes\sigma_z}{\Psi_j} =
\delta_{ij}$ are properly normalizable. We will often work directly in the
$\Psi_0$ and $\Psi_1$ basis, writing $\ket 0 = (1,0)$ and $\ket 1=(0,1)$ as in
Fig.~\ref{fig:bloch-sphere}. The use of braket notation here is
  a formalism of convenience arising from the close mathematical similiarities
  between our system and single-particle quantum mechanics. However, we
  emphasize early on that this is a purely notational convenience; it is
  impossible to realize many-body quantum phenomena, such as entanglement, in a
  purely semiclassical magnonic system.

Since $\cosh x > \sinh x$ for all real $x$, the magnitude of the spin wave
precession is clearly dominated by the $A$ sublattice in $\Psi_0$ and the $B$
sublattice in $\Psi_1$. One can see from Fig.~\ref{fig:lr} that these two modes
carry opposite magnetization, since the $\hat S_z$ component of the sublattices
must differ if one of the sublattices dominates. The reduction of magnetization
on each sublattice is simply given by the squared magnitude of the lattice spin
wave, so that the total magnetization induced by a spin wave is
\begin{equation}
  m_z = -S\mel{\Psi}{(\bbid_2\otimes\sigma_z)}{\Psi} \;,
\end{equation}
which will be negative for right-handed waves proportional to $\Psi_0$ and
positive for left-handed waves proportional to $\Psi_1$. This operator
$\bbid_2\otimes\sigma_z$ corresponds to a so-called non-geometric
symmetry.~\cite{proskurin:2017aa} It has sometimes been given as the
\emph{definition} of spin wave chirality. In electromagnetic analogies for AFM
spin wave dynamics it corresponds to optical helicity,~\cite{proskurin:2017aa}
where the corresponding conserved quantity is the so-called
zilch.~\cite{cameron:2012aa}

So far, we have dealt only with a block diagonal Hamiltonian. Restricting to the
positive energy subspace, we see that $\mathcal H$ has no off-diagonal terms
that connect $\Psi_0$ and $\Psi_1$. If such terms existed, we could manipulate
the total spin carried by the spin wave in-transit, rotating our spin wave state
within the degenerate eigensubspace. We may imagine that the coefficients
balancing these eigenvectors in a superposition $\ket{\bEta} = \eta_0\ket{0} +
\eta_1 \ket{1}$ define a new degree of freedom which we refer to as the magnonic
isospin. \emph{The desire to exploit this internal degree of freedom motivates
  the remainder of the paper.}

\subsection{Spin texture, characteristic length scales, and perturbative
  parameters}
\label{sec:spin-texture-maintext}

In order to control $\bEta$, we must find a way to break the degeneracy between
the right- and left-handed modes; that is, we must break whatever symmetries are
protecting either conservation of chirality (that is, the block-diagonality of
$\mathcal H$) or conservation of the relative phase between right- and
left-handed modes. In this paper, the main tools we consider for this purpose
are spin texture and the Dzyaloshinskii-Moriya interaction (DMI). The latter is
well-known and we introduce the approrpiate free energies when they are needed.
Spin texture, however, is somewhat more subtle, so we briefly review theoretical
tools for handling it. These techniques have been used to great success in
describing transport effects arising from both
ferromagnetic\cite{Dugaev:2005aa,Guslienko:2010kq,Nakata:2016wu,vanHoogdalem:2013:texture}
and antiferromagnetic\cite{Cheng:2012kl,kouki:2017:magnonTI,Buhl:2017wf}
textures.

To describe the spin texture in our formalism, we encode the texture in a
rotation matrix $R$ defined by $R \bn = |\bn|\hat z$. This rotation matrix
induces a generator of infinitesimal spin rotations, $(\partial_\mu R)R^T$,
which itself can be regarded as a collection of vector potentials $A_\mu^x J_x +
A_\mu^y J_y + A_\mu^z J_z = (\partial_\mu R)R^T$, the decomposition being
directed through the standard generators\footnote{We take these to be the
  real-valued generators, so that the rotation by $\theta$ about $\hat{e}_j$
  given through the exponential map $\exp(\theta J_j)$. Some authors include an
  $i$ in this expression for the rotation matrix in order to make it look
  manifestly ``unitary'', and those authors must also have imaginary-valued
  $J_j$ matrices to compensate.} of 3D rotations $J_j$. Here $\mu$ is a
spacetime index, and the components $A_\mu^j$ define the (1+d)-vectors $A^j =
(A^j_t, \bA^j)$.

Because our spin texture is described with respect to the $\hat{z}$-axis,
$\bA^z$ will be of paramount importance. It gives rise to an emergent magnetic
field $\bB = \nabla \times \bA^z$ that produces a Lorentz force on magnons in
Eqs.~\eqref{eq:sc-eoms}, and the temporal component $A_t^z$ likewise produces an
emergent electric field. We will usually describe the influence of the other two
potentials through the complex variable $\mathcal A_\mu = (A^x_\mu + i
A^y_\mu)/\sqrt 2.$ For more information on these fields, the reader is referred
to Appendix \ref{sec:spin-waves-spin}. For a full discussion of this gauge field
formalism in the treatment of spin texture, the reader may check
Refs.~\onlinecite{Dugaev:2005aa, Guslienko:2010kq}.

We will soon need an approximation scheme to deal with the many perturbative
effects---anisotropy, DMI, \emph{et cetera}---of our spin wave system. Since
$A_\mu^j$ is a derivative of the texture-defining angles, let it define a
characteristic length scale $\lambda$ of the system,
\begin{equation}
  |A_\mu^j| \sim \frac{1}{\lambda}.
\end{equation}
In textured systems with DMI, the characteristic length scale is
proportional~\cite{Rohart:2013ef} to $J/D$, where $D$ is the DMI
strength\footnote{Depending on Moriya's rules~\cite{Moriya:1960go} for how a
  system breaks inversion symmetry, the vector structure of the free energy term
  may differ, but will in general descend from some lattice Hamiltonian
  $\sum_{\langle i j\rangle} \bD_{ij}\cdot(\ma^i\times\mb^j)$.} $\mathcal
F_\text{DMI} = D\ma \cdot (\nabla \times \mb)$. Therefore \footnote{This length
  scale is set when $D$ is large enough to influence the spin texture; the
  critical value of $D$ is typically set by anisotropy. So if $D$ is large
  enough to influence the texture, then it is order $|\covA{}|$, and otherwise
  we are harmlessly overestimating the importance of contributions from $D$.}
\begin{equation}
  D/J \sim |A_\mu^j|.
\end{equation}
In systems with easy axis anisotropy, meanwhile, the well-known characteristic
length of a domain wall is $\sqrt{J/K}$, and thus
\begin{equation}
  K/J \sim |A_\mu^j|^2.
\end{equation}
Finally, the local magnetization\footnote{Note that $R\bmag$ has no $\hat z$
  component, since $\bmag \perp \bn$ by construction.} $\mu = (R\bmag)\cdot(\hat
x + i\hat y) / \sqrt 2$ scales as a derivative of the staggered
order,~\cite{Tveten:2015aa}
\begin{equation}
  \mu \sim |A_\mu^j|.
\end{equation}
As it happens, the magnetization will, in our calculations, never show up as a
lone linear-order term; even so, the quadratic terms $O(\mu^2) = O(K/J)$ must be
preserved.

We have established a hierarchy of perturbative orders based on a single
parameter, $|A|$. In our spin wave treatment, we will keep terms up to order
$\partial A \sim O(A^2)$, that is, to linear order in the emergent
electromagnetic field $\bB = \nabla\times\bA^z$.

\subsection{Matrix structure of the spin wave Hamiltonian}

Once we add extra terms to the free energy---spin texture, the
Dzyaloshinskii-Moriya interaction, and so on---the equation of motion becomes
\begin{equation}
  i  (\tau_z\otimes\sigma_z) \dot\Psi = \frac{\epsilon^d}{n S}\frac{\delta F}{\delta \bar\Psi}- A_t^z(\bbid_2 \otimes\sigma_z)\Psi
  \label{eq:schro-4-general}
\end{equation}
so that the spin wave Hamiltonian is given through $\mathcal H \Psi = \delta
F/\delta\bar\Psi$. Here, $\epsilon$ is the lattice constant, $d$ is the
dimensionality of the lattice, and $n = 1 + |\mu|^2$ is the effective index of
refraction for the spin wave speed, when viewed from the perspective of the wave
equation governing staggered order dynamics. Eq.~\eqref{eq:schro-4-general}
prescribes the correct harmonic spin wave theory for any free energy $F$, where
the basis vectors $\{\alpha, \beta, \alpha^*,\beta^*\}$ are now defined as the
\emph{purely in-plane} fluctuations of the sublattice spin wave modes after the
active rotation by $R$ of the ground state texture. The detailed derivation of
Eq.~\eqref{eq:schro-4-general} is given in Appendix~\ref{sec:quasistatic}.

For concreteness, we now present the detailed matrix form of the exchange
interaction Hamiltonian. Beginning from Eq.~\eqref{eq:free-energy-exch}, we
rotate the fields by $R$ and change variables to the in-plane complex
fluctuations $\alpha$, $\beta$, $\alpha^*$, and $\beta^*$. The corresponding
Hamiltonian for the homogeneous exchange interaction is
\begin{equation}
  \mathcal H_\text{hom} = \frac{\calZ}{2}\left(\begin{matrix}
      1 - 3|\mu|^2 & 1 - |\mu|^2 & \mu^2 & -\mu^2 \\
      1 - |\mu|^2 & 1 - 3|\mu|^2 & -\mu^2 & \mu^2 \\
      \bar\mu^2 & -\bar\mu^2 & 1 - 3|\mu|^2 & 1 - |\mu|^2 \\
      -\bar\mu^2 & \bar\mu^2 & 1 - |\mu|^2 & 1 - 3|\mu|^2 \\
    \end{matrix}\right)
\end{equation}
where the bar over $\bar\mu$ (and, later, over $\bar\calA$) indicates complex
conjugation. The inhomogeneous exchange interaction $\mathcal H_\text{inhom}$,
meanwhile, is given by $\calJ/(2n)$ times the matrix
\begin{widetext}
  \begin{equation}
    \left(\begin{matrix}
        -2|\calA|^2 & (\nabla - i\bA^z)^2 + \bm{\Delta}\cdot\nabla - |\calA|^2 & 0 & -(\mu\nabla)^2 + 4i\mu\calA\cdot\nabla + \calA^2\\
        (\nabla - i\bA^z)^2 - \bm{\Delta}\cdot\nabla - |\calA|^2 & -2|\calA|^2 & -(\mu\nabla)^2 - 4i\mu\calA\cdot\nabla + \calA^2 & 0 \\
        0 & -(\bar\mu\nabla)^2 + 4i\bar\mu\bar\calA\cdot\nabla + \bar\calA^2 & -2|\calA|^2 & (\nabla + i\bA^z)^2 + \bm{\Delta}\cdot\nabla - |\calA|^2 \\
        -(\bar\mu\nabla)^2 - 4i\bar\mu\bar\calA\cdot\nabla + \bar\calA^2 & 0 & (\nabla + i\bA^z)^2 - \bm{\Delta}\cdot\nabla - |\calA|^2 & -2|\calA|^2  \\
      \end{matrix}\right)
  \end{equation}
\end{widetext}
where $\bm{\Delta} = 2i(\bar\mu\calA - \mu\bar{\calA})$. These matrix
Hamiltonians, and the Hamiltonians corresponding to any other 2-site
interaction, exhibit notable structural differences when the synthetic AFM case
is considered instead. In the Supplemental Material, we have provided a
\emph{Mathematica} notebook that automates the derivation of $\mathcal H$ for
any free energy given in terms of $\ma$ and $\mb$. It also contains pre-computed
Hamiltonians for anisotropy, DMI, external fields, and so on, which we use in
our applied examples later in the paper.

\section{Non-abelian wavepacket theory}
\label{sec:non-abel-wavep}
\begin{figure*}
  \centering
  \includegraphics[width=\textwidth]{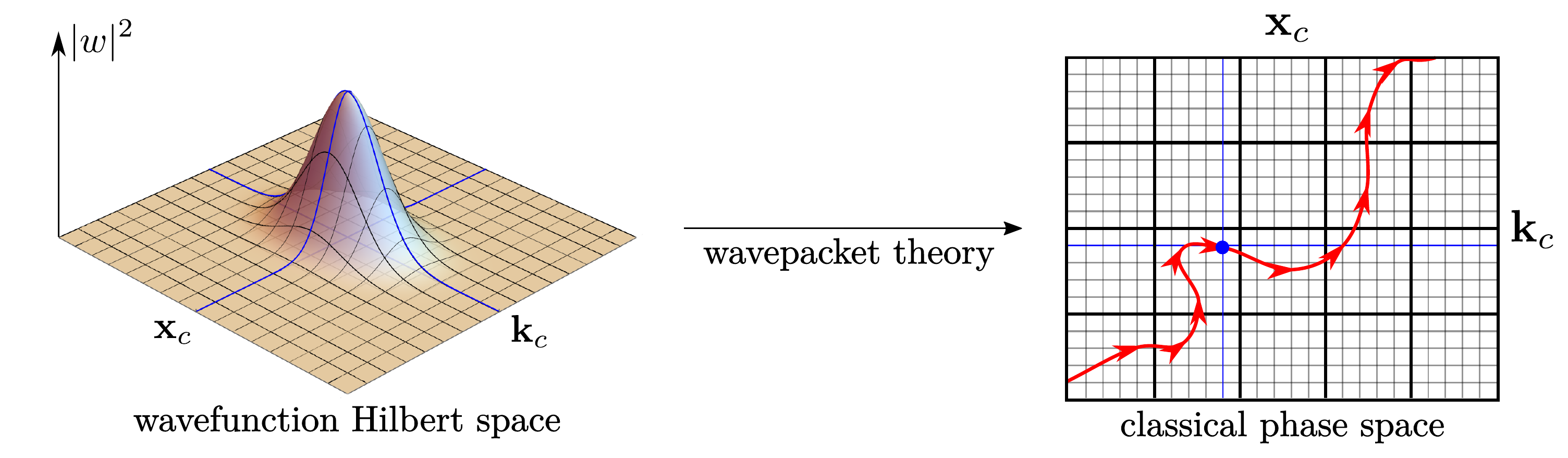}
  \caption{Under the assumptions of wavepacket theory, the magnon wavepacket has
    its magnitude $w(\bq,t)$ strongly localized in real and momentum space.
    Consequently, the wavefunction is sufficiently specified by its mean
    coordinates $(\bxc, \bkc)$ on phase space. The wavepacket theory machinery
    uses this assumption to resolve the wave theory (left) described by
    Eq.~\eqref{eq:schro-4-general} into a particle theory (right) described by
    the classical Lagrangian Eqs.~\eqref{eq:collected-wp-lagrangians}. Not
    pictured is the isospin degree of freedom, which lives in an SU(2) fiber
    over the classical phase space. The full semiclassical dynamics described by
    Eqs.~\eqref{eq:sc-eoms} occurs on the induced fiber bundle.}
  \label{fig:phase-spaces}
\end{figure*}
In this section, motivated by the need to derive $\bEta$-dynamics from
Eq.~\eqref{eq:schro-4-general} in the case of spatial inhomogeneity, we apply
the machinery of non-abelian wavepacket theory. ~\cite{culcer:2005coherent} What
we call ``wavepacket theory'' was originally developed in a Letter by Chang and
Niu ~\cite{Chang:1995zz} to explain the Hofstadter butterfly spectrum, after
which their treatment was codified by Ref.~\onlinecite{Sundaram:1999ht}. Since
then, the theory has been applied in a variety of contexts, sometimes requiring
extensions of the theory to account for unique features of a particular physical
problem.~\cite{Zhang:2006bj,DiXiao:2009kw,Xiao:2015tz}

The most relevant extension for our purposes---and indeed, one of the most
ambitious and interesting developments in wavepacket theory---is the treatment
of multiple degenerate bands.~~\cite{culcer:2005coherent,shindou:2005aa} In this
case, the theory is called \emph{non-abelian wavepacket theory} because, in
dealing with a vector of multiple band energies at once, the ``coefficients''
must become matrix valued (and therefore, generally speaking, an element of a
non-abelian matrix representation) in order to act on the multi-band
wavefunction. In this paper, we extend the non-abelian wavepacket theory to
account for both the unusual $\tau_z\otimes\sigma_z$ factor in our Lagrangian
and our explicitly \emph{a priori} non-abelian gauge
field.\footnote{Ref.~\onlinecite{shindou:2005aa} deals with a non-abelian gauge
  field in the non-abelian wavepacket theory, but only to derive it implicitly
  from the band structure; that is, they do not write down such a gauge field in
  their starting Lagrangian, but instead tease it out. Our approach is more
  similar to the original technique by Ref.~\onlinecite{Chang:1996zz}, but
  lifted to the non-abelian case (and of course to the non-Euclidean Hilbert
  space). These two approaches should ultimately agree, though each is
  operationally preferable for certain classes of problems.} A detailed
derivation involving the internal workings of wavepacket theory are crucial for
establishing our main results. Since details of wavepacket theory, even in the
abelian case, are not widely studied, we carefully guide the interested reader
through the derivation in Appendix \ref{sec:nonab-wavep-theory}.

The basic idea of abelian wavepacket theory is to consider a momentum-space
superposition $|W\rangle = \int w_q \psi_q \dd[d]\bq$ of eigenvectors, where the
eigenvectors are drawn from the spectrum of the Hamiltonian evaluated at some
$(\bxc,\bqc)$ on a classical phase space. In non-abelian wavepacket theory, the
eigenvector is expressed as a general state lying in the degenerate subspace
spanned by our right- and left-handed modes,
\begin{align}
  \ket{W(\bxc,\bkc)} &= \int \dbq w(\bq,t)\big[\nonumber\\
  \eta_0(\bq,t) &\ket{\Psi_0(\bq,t)}
                  + \eta_1(\bq,t) \ket{\Psi_1(\bq,t)}\big].
                  \label{eq:gauged-wp}
\end{align}
The coefficient $w$ gives the shape of the wavepacket, as in
Fig.~\ref{fig:phase-spaces}. The vector $\ket\bEta = (\eta_0,\eta_1)$ is, again,
called the \emph{isospin}. We demand that the otherwise generic wavepacket
possess
\begin{enumerate}
  \label{enum:wp-defining-props}
\item a momentum space distribution localized enough to be approximated as
  $\delta(\bq-\bqc)$,
\item a well-defined mean position $\bxc = \mel{W}{\hat\bx}{W}$, and
\item sufficient spatial localization that the environment where the wavepacket
  has appreciable support is approximately translationally invariant.
\end{enumerate}
These assumptions form a set of sufficient conditions under which a
wavefunction's semiclassical dynamics can be formulated, using wavepacket
theory, on a classical phase space $\Gamma \ni (\bxc,\bqc)$. The non-abelian
version, Eq.~\eqref{eq:gauged-wp}, includes an $\bEta$-valued fiber over
$\Gamma$.

By appealing to the time-dependent variational principle, we can write down the
Lagrangian which generates the equation of motion \eqref{eq:schro-4-general},
namely $L_\text{WP} = \mel{\Psi}{\mathcal L}{\Psi}$ with
\begin{equation}
  \label{eq:wp-lagrangian-density}
  \mathcal L = i(\tau_z\otimes\sigma_z)\frac{\dd}{\dd{t}} - \mathcal H - A_t^z\sigma_z.
\end{equation}
We then assume $\ket W$ as the solution for $\ket\Psi$. Since the wavepacket is
sufficiently\footnote{As long as the wavepacket satisfies the assumptions listed
  in the main text, the functional form of the wavepacket does not affect the
  outcome of wavepacket theory.} described by the 3-tuple $(\bxc,\bqc,\bEta)$,
we can reduce $L_\text{WP}$ to a Lagrangian of the phase space variables $\bxc$,
$\bqc$, and $\bEta$ that specify $\ket W$. The result is
\begin{subequations}
  \label{eq:collected-wp-lagrangians}
  \begin{align}
    L_\text{WP} &= L_{\dd t} + L_H + L_\text{EM},\;\text{where}\\
    L_{\dd t} &= \mel{\tilde\bEta}{\dot\bx_c\cdot\fraka_\bx
                +  \dot\bq_c\cdot\fraka_\bq
                + \fraka_t 
                + i\partial_t}{\tilde\bEta} - \dot\bq_c\cdot\bxc,\\
    L_H &=- \bra{\bEta} {\mathscr H} \ket{\bEta},\quad\text{and}\\
    L_\text{EM} &=-\dot\bA^z\cdot\Gamma_{\bq}-\chi(\dot\bA^z\cdot \bxc + A^z_t).\label{eq:L-A-subeq}
  \end{align}
\end{subequations}
$L_{\dd{t}}$, $L_H$, and $L_\text{EM}$ derive from the time derivative,
Hamiltonian, and emergent field terms from Eq.~\eqref{eq:wp-lagrangian-density},
respectively. Here in the main text, we simply pause to describe the various
physical variables in Eqs.~\eqref{eq:collected-wp-lagrangians} that fall out of
the derivation.

First, let us define the $4\times 2$ matrix
\begin{equation}
  E = \ket{0} \bra{\Psi_0} + \ket{1} \bra{\Psi_1}.
\end{equation}
where $\ket 0$ and $\ket 1$ are understood as the basis vectors $(1,0)$ and
$(0,1)$ for the isospin $\ket{\bEta} = \eta_0 \ket 0 + \eta_1 \ket 1$. $E$ is
essentially a change of basis matrix (which chooses $\Psi_0$ and $\Psi_1$ as the
canonical basis vectors), followed by a projection to the forward-time
degenerate Hilbert subspace that they span. $E^\dagger$ represents the embedding
of the isospin dynamics into the full spin wave dynamics, and as such the
induced isospin Hamiltonian is given by
\begin{equation}
  \label{eq:isospin-h-by-embedding}
  \mathscr H = E\mathcal H E^\dagger.
\end{equation}
Next, we define the various $2\times 2$ matrices $\hat a_\mu$. These are the
matrix-valued Berry connections in isospin space,
\begin{equation}
  \fraka_\mu^{ij} = \langle\Psi^i_\bq|i\sigma_z\partial_\mu \Psi^j_q\rangle.
\end{equation}
These diagonal matrices will generate Berry curvatures---effective, emergent
magnetic fields---in the equations of motion.~\cite{culcer:2005coherent} The
term $\Gamma_\bq = \mel{\bEta}{\tau_z\hat a_\bq}{\bEta} -
\mel{\bEta}{\tau_z}{\bEta}\mel{\bEta}{\hat a_\bq}{\bEta}$ arises uniquely due to
the $\tau_z\otimes\sigma_z$ metric structure of our full four dimensional
Hilbert space, and is absent from existing non-abelian wavepacket theories which
deal only with Euclidean spaces. It gives rise to a nonlinear potential $V_\chi
= \delta \Gamma_\bq/\delta \bEta$. Finally, the tilde decoration on $\tilde\bEta
= \calG\bEta$ refers to a gauge transformation $\calG =
\exp(-i(\tau_z\otimes\bbid_2)\bA^z\cdot\bx)$ discussed in
Eq.~\eqref{eq:G-gauge-xform}. Hamilton's principle $\delta S = 0$ gives us
equations of motion for the dynamical variables:
\begin{subequations}
  \label{eq:sc-eoms}
  \begin{align}
    \dot{\bq}_c &= \chi\left(\bE + \dot\bx_c\times\bB\right)-\frac{\partial \mathcal E}{\partial \bx_c}, \\
    \dot\bx_c &=\frac{\partial\mathcal E}{\partial \bq_c} + \langle\bOmega^{\bq\bq}\rangle\dot\bq_c + \langle\bOmega^{\bq\bx}\rangle\dot\bx_c + \langle\Omega^{\bq t}\rangle,\;\text{and} \\
    i\frac{\dd}{\dd t}\bEta
                &= \left[\mathscr{H} - \mathscr A_t + \tau_z A^z_t + \hat{V}_\chi\right]\bEta,\label{eq:isospin-schrodinger}
  \end{align}
\end{subequations}
with $\mathscr A_t = \dot\bx_c\cdot\hat a_\bx + \dot\bq_c \cdot \hat a _\bq +
\hat a _t$, $\mathcal E$ the linearly perturbed spin wave energy (as in
Ref.~\onlinecite{culcer:2005coherent}), and $\bOmega$ are the various Berry
curvature terms,
\begin{equation}
  \langle \Omega^{\alpha\beta}_{\mu\nu}\rangle = \mel**{\bEta}{\left(\frac{\partial \fraka_{\beta^\nu}}{\partial \alpha^\mu} - \frac{\partial \fraka_{\alpha^\mu}}{\partial \beta^\nu}\right)}{\bEta}.
\end{equation}
Finally, the emergent electromagnetic fields are $\bB = \nabla\times \bA^z$ and
$\bE = \nabla A_t^z$, familiar to those who have studied magnetic skyrmion
physics.~\cite{Nagaosa:2013aa, Guslienko:2010kq}

The reduction of $L_\text{WP}$ to single-particle Lagrangian
\eqref{eq:collected-wp-lagrangians} is quite technical, and we relegate the
derivation to Appendix \ref{sec:nonab-wavep-theory}. The process is illustrated
schematically in Fig.~\ref{fig:phase-spaces}. The equations of motion
Eqs.~\eqref{eq:sc-eoms}, as well as their derivation, are tightly related to the
results of Ref.~\onlinecite{culcer:2005coherent}. The differences arise due to
the non-Euclidean metric $\tau_z\otimes\sigma_z$ in the Lagrangian. This new
geometry gives rise to the dynamical charge $\chi =
\bra{\bEta}\tau_z\ket{\bEta}$ coupled to the Lorentz force, and also gives rise
to the nonlinear potential $V_\chi$ (through $\Gamma_\bq$).

Though $V_\chi$ can contribute at $O(\covA{}^2)$ in perturbation theory in
principle, it only contributes at third order or above for the interactions we
consider concretely in this article. To contribute in our formalism, it would
require that $\fraka_\bq$ manifest at leading order in the perturbation
theory---or else that we go to higher order in the perturbation theory, as a
non-abelian and non-Euclidean extension of second-order wavepacket
theory.~\cite{yang:2014aa, yang:2015aa} If such a system could be identified,
then the physics of $V_\chi$---which induces a Gross-Pitaevskii equation for the
isospin---could be quite interesting. In the coupling between a wavepacket and a
rigid soliton, for instance, we see that this term produces at leading order a
force proportional to $\dot\chi$. Thus, a change in the spin carried by the
magnon produces a real-space force on the soliton. We leave the search for
systems in which $V_\chi$ could produce significant effects to future research.

Finally, let us caution the reader that Eq.~\eqref{eq:isospin-schrodinger} gives
the dynamics of the isospin, which is defined with respect to the $A$ and $B$
sublattices---not with the laboratory frame. A right-handed mode, for instance,
is by our definition always dominated by the $A$ sublattice---which means that
it carries opposite spin on either side of a domain wall. To return to the lab
frame, one should apply the inverse rotation operator $R^{-1}$ to the spin
texture. To extract the lab-frame spin, then, lift $R^{-1}$ to SU(2) by the
standard homomorphism~\cite{Tinkham:2010ty} and apply it to the isospin. The
(semiclassical) spin carried at time $t$ by the magnon with isospin $\bEta(t)$
is then\footnote{As usual, there are two distinct SU(2) rotations corresponding
  to the O(3) rotation $R^{-1}$ of the ground state. They differ by a sign, but
  this sign is immaterial in calculations of observable quantities such as
  $m_z$.}
\begin{equation}
  \ket{\bm{s}(t)} = \left(\begin{matrix}
      e^{i(\psi + \phi)}\cos\frac{\theta}{2} & e^{i(\psi-\phi)}\sin\frac{\theta}{2}\\
      -e^{-i(\psi - \phi)}\sin\frac{\theta}{2} & e^{-i(\psi+\phi)}\cos\frac{\theta}{2}
    \end{matrix}\right) \ket{\bEta(t)} \label{eq:su2-rot}
\end{equation}
where the Euler angles defining the texture are evaluated at $\bx_c(t)$. The
observable magnetization carried by the isospin is then
$m_z=-\mel{\bm{s}}{\sigma_z}{\bm{s}}$, the sign arising from the fact that
right-handed waves $\ket 0$ carry negative spin. Since we are generally
interested in systems with easy-axis anisotropy, the matrix transformation in
Eq.~\eqref{eq:su2-rot} will typically result in a simple sign $m_z =
\mp\mel{\bEta}{\sigma_z}{\bEta}$ depending on whether the local \Neel/ order
along the easy axis is pointing along $\pm \hat z$.

The key result of our wavepacket analysis, as regards the remainder of this
article, is that the isospin $\bEta$ obeys an emergent \Schrodinger/ equation,
and its dynamics are therefore governed by unitary time evolution. By tailoring
our Hamiltonian, we can generate unitary rotations about multiple different axes
in isospin space. We display a collection of different rotations in the coming
examples, which taken together will be sufficient to generate any generic
rotation (in three Euler angles) of the isospin.

\section{Application to selected magnonic primitives}
\label{sec:appl-select-magn}
In the previous section, we derived a set of semiclassical equations governing
the isospin-coupled dynamics of a magnon wavepacket. Now we apply that formalism
to two AFM magnonic systems: a gated 1D wire, and a 1D domain wall. We conclude
by mentioning the effects of magnetic fields and hard-axis anisotropy.

\subsection{A gated AFM nanostrip: the magnon FET}
\label{sec:magfet}
\begin{figure*}
  \centering \includegraphics[width=\textwidth]{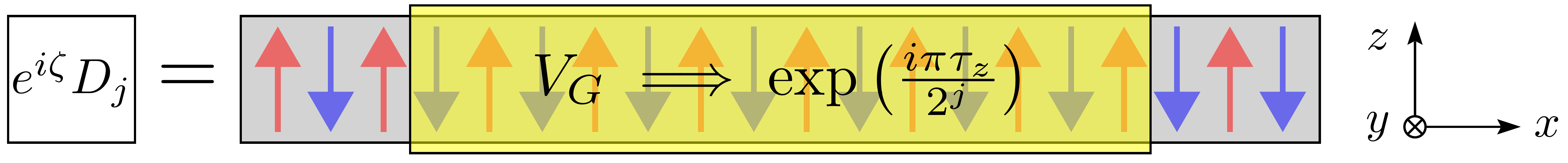}
  \caption{The system under investigation in Sec.~\ref{sec:magfet}. An in-plane
    easy-axis ($\hat z$) AFM is oriented in a nanostrip geometry, perpendicular
    to the easy-axis ($\hat x$). A section of the sample is subjected to a gate
    voltage $V_G$ applied normal to the sample plane, in the $\hat y$-direction.
    We show in Eq.~\eqref{eq:dmi-ds-h} that the resulting isospin dynamics
    corresponds to a rotation about $\sigma_z$ on the Bloch sphere
    Fig.~\ref{fig:bloch-sphere}. We define here the notation $D_j =
    \text{diag}(1, e^{i\pi/2^j})$, and is given by this applied DMI gate up to a
    dynamical phase $e^{i\zeta}$. Note, as a reference, that $\tau_z = D_0$.}
  \label{fig:dmigate}
\end{figure*}

In this section, we consider the application of a gate voltage across a 1D AFM
nanowire (extended along $\hat x$) with in-plane easy-axis anisotropy (along
$\hat z$). Our motivation is threefold. First, this gate will be extremely
important in our device proposals later in the article, so it is worthwhile to
present the theoretical treatment here. Second, this simple example which does
\emph{not} possess any spin texture will provide a transparent presentation to
demonstrate the general solution method to the reader. Finally, solving this
problem---which has already been considered in the \Neel/ vector picture, for
the special case of linearly polarized waves, by
Ref.~\onlinecite{Cheng:2016kv}---will serve as a validation of our theoretical
methods against the literature.

The free energy has four parts: homogeneous and inhomogeneous exchange,
easy-axis anisotropy, and DMI. The first three of these are the same as is given
in Eqs.~\eqref{eq:all}, and the DMI term is
\begin{equation}
  F_\text{DMI} = \frac{1}{2}\int \bD \cdot \left[  \ma \times \partial_x \mb + \mb \times \partial_x \ma\right] \dd x,
\end{equation}
where $\bD = D \hat z$. From the corresponding $4 \times 4$ Hamiltonian, we
construct the $2\times 2$ isospin Hamiltonian by using the embedding $E^\dagger$
and Eq.~\eqref{eq:isospin-h-by-embedding}. Writing out $\mathscr H = \mathscr
H_0 + \mathscr H_j \sigma_j$ explicitly for this problem, we find that it has an
unimportant\footnote{The constant ($\bbid_2$) part is familiarly unimportant in
  a truly 1D system, but with multibranch devices such as the one in
  Fig.~\ref{fig:nalc} it introduces a \emph{very much important} relative U(1)
  phase between different spin wave channels.} constant part as well as a
$\sigma_z$ component:
\begin{equation}
  \mathscr H_z = \frac{J|D|k\epsilon\left(1 - \frac{(k\epsilon)^2}{2}\right)}{\hbar s \sqrt{2KJ + (Jk\epsilon)^2}}.
  \label{eq:hz-mfet}
\end{equation}
If we assume both that $K/J$ is small and that $k$ is in a regime where the
distance between the split bands is constant in $k$, namely well above the
resonance frequency, then the denominator of Eq.~\eqref{eq:hz-mfet} can be
approximated merely by $\hbar s J k \epsilon$, canceling the linear contribution
in the numerator and leaving only the constant term with a weak quadratic
correction. Making these approximations in Eq.~\eqref{eq:hz-mfet}, we arrive at
an isospin hamiltonian
\begin{equation}
  \label{eq:dmi-ds-h}
  \mathscr H_z = D/S.
\end{equation}
How does this Hamiltonian act on the isospin state? Since we are dealing with a
\Schrodinger/ equation (Eq.~\eqref{eq:isospin-schrodinger}), we need only
compute the unitary time evolution operator
\begin{align}
  U(t_1,t_0) &= \exp\left[\frac{i\tau_z}{\hbar s}\int_{t_0}^{t_1} D \dd{t} \right]\\
             &= \exp\left[\frac{i\tau_z}{\hbar s}\left(\pdv{\omega}{k}\right)^{-1}\int_{x_0}^{x_1} D \dd{x}\right].
\end{align}
This is a rotation operator in isospin space, rotating about the $\hat z$-axis
on the Bloch sphere by a total angle proportional to $D$ and the length of the
gate, but inversely proportional to the spin wave speed $\partial_k\omega$ and
the spin magnitude $S$. The rate of rotation on the Bloch sphere works out to
\begin{equation}
  \nabla\phi = \frac{1}{s}\cdot\frac{D}{J},
\end{equation}
Note that we have cited the rate of rotation on the Bloch sphere, where $\phi$
is the azimuthal angle---this differs by a factor of two from the polarization
angle of the staggered order. $X$- and $Y$-polarized states, which appear to be
rotations of $\pi/2$ away from each other in the trace of a spin excitation, are
actually $\pi$ away from each other on the Bloch sphere
(Fig.~\ref{fig:bloch-sphere}).

Because the rate of rotation scales with the DMI itself, the rotation on a
single gate can be manipulated on-line simply by modulating the gate voltage. We
concerned ourselves in Ref.~\onlinecite{Cheng:2016kv} (with which our result in
Eq.~\eqref{eq:dmi-ds-h} agrees) mostly with a rotation between $X$ and $Y$
polarizations, but access to generic rotations will be crucial
  for a mature implementation of nonabelian magnonics.

\begin{figure*}
  \centering \includegraphics[width=\textwidth]{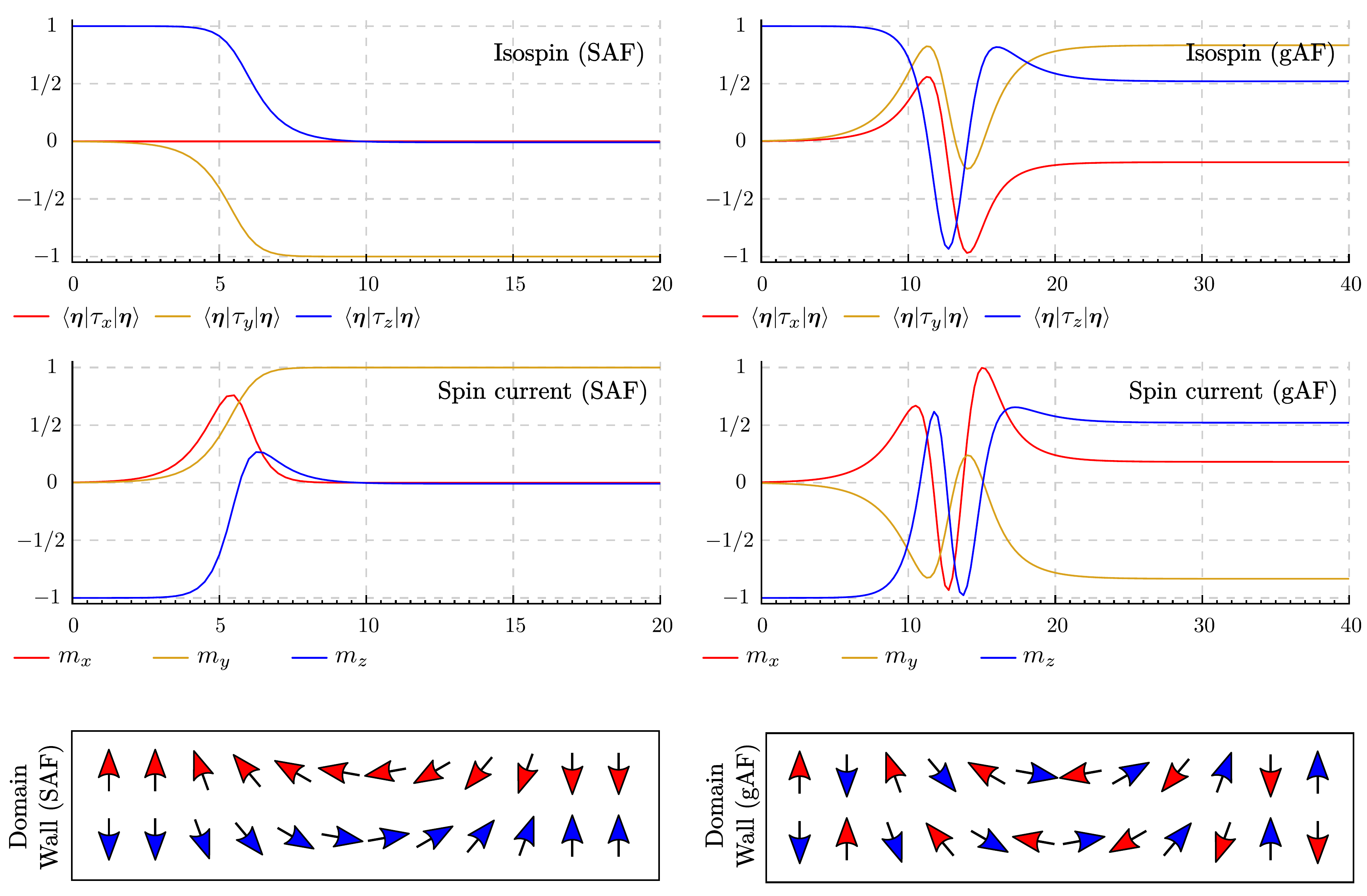}
  \caption[Single wavepacket]{Semiclassical dynamics of a single magnon passing
    through a Bloch-type domain wall. The horizontal axes represent time, given
    in picoseconds. \emph{Left}: integration of Eqs.~\eqref{eq:sc-eoms} for a
    wavepacket, initially with right-handed polarization $\bEta = \ket 0$,
    passing through a domain wall in a synthetic AFM. The SAF material
    parameters were taken from YIG, and the initial frequency of the wavepacket
    was tuned to result in a $\pi/2$ rotation on the Bloch sphere. The top plot
    gives the isospin expectation values; bottom, these have been rotated to
    give the true spin current. \emph{Right}: the same semiclassical dynamics,
    domain wall, and YIG parameters are simulated, but the system is assumed to
    be $g$-type AFM. We merely substitute the ferromagnetic exchange for the
    inhomogeneous exchange, and antiferromagnetic for homogeneous exchange.
    Because $\mathcal T\mathcal I$ symmetry is broken in the $g$-type
    configuration, the rotation is unavoidably more complex. \emph{Bottom}:
    schematic illustration of a $g$-type versus a synthetic AFM domain wall. We
    have illustrated \Neel/ type walls for simplicity, but the calculation was
    done for Bloch-type walls.}
  \label{fig:single-wp-retarder}
\end{figure*}
\subsection{Domain wall retarder}
\label{sec:application-dw}

Since applied AFM magnonics has become fashionable in the last decade,
the AFM domain wall has undergone quite a bit of new analysis
~\cite{lan:2017aa,Tveten:2014aa,Kim:2014aa,Kim:2015el}, and in decades
past was a prototypical nontriviality for the AFM nonlinear sigma
model.~\cite{Papanicolaou:1995iy,Ivanov:1995bf,Manton:vEuV_4gy} Many
such studies have concluded that spin waves passing through a domain
wall experiences a relative frequency shift between the right- and
left-handed components.~\cite{Kim:2014aa} In systems with DMI, they can
express even more pronounced shifts between linearly polarized modes,
giving rise to a retarding waveplate effect.~\cite{lan:2017aa} Our formalism
allows us to calculate this shift precisely, and in terms of the SU(2)
isospin.

In this section, we consider a Bloch-type domain wall in a synthetic
AFM with easy-axis anisotropy and a bulk-type DMI.  Take the Walker
solution for the 1D texture as
\begin{equation}
  \theta(x) = -2\arctan\left(\exp\frac{x}{\lambda}\right) \quad\text{and}\quad \phi(x) = -\pi/2,
\end{equation}
with $\lambda = \sqrt{J/K} = O(\covA{}^{-1})$ the domain wall width.

With this texture, we can immediately calculate the texture-induced
gauge fields from Eq.~\eqref{eq:explicit-A-vectors} (taking $\psi = 0$
for concreteness): we have $\bA^z = 0$ and
\begin{subequations}
\begin{align}
  \bA^x &= \frac{1}{\lambda}\sin\psi \sech \frac{x}{\lambda}\\
  \text{and}\quad \bA^y &= \frac{1}{\lambda}\cos\psi\sech \frac{x}{\lambda}\\
  \implies \mathcal A &= \frac{i}{\lambda \sqrt 2}\sech\frac{x}{\lambda},
\end{align}
\end{subequations}
where we have suppressed the spacetime index since there is only
one.\footnote{There is no temporal index because the DW is assumed to
  be static.} The bulk-type DMI is written as $\bD_{ij} = D\hat{r}_{ij}$, and
minimization of DMI energy has been used to determine $\phi(x)$.   

Using spin wave Hamiltonian $\mathcal H$ for synthetic AFMs detailed
in the Supplementary Material, we compute the appropriate coefficients
of the semiclassical dynamics in Eqs.~\eqref{eq:sc-eoms}. The
resulting isospin Hamiltonian has an again unimportant $\bbid_2$
component as well as a $\tau_x$ component. The $\tau_x$ term is
\begin{equation}
  \mathscr H_x = \frac{\mathcal DK (\calZ  + 2\calJ k^2)}{4\ell \sqrt{\calJ K }}\sech\frac{x}{\lambda}.
\end{equation}
Since $\mathscr H$ has no other nontrivial component, we see immediately that it
will carry out a rotation of the isospin about $\hat x$ on the Bloch sphere, and
will do so most strongly near the center of the domain wall due to the
exponential localization provided by $\sech(x/\lambda)$.

From there, we have $\mathcal E$ (since we have $\mathcal H$),
$\mathscr H$ (since we have $\mathcal H$ and $E$), and we know that
the $\bB = \bE = 0$ by inspection of $\bA^z$. The other Berry
curvature terms are easily seen to vanish as well. We immediately
construct the semiclassical equations Eqs.~\eqref{eq:sc-eoms} and
integrate them with an adaptive-step size Runge-Kutta-Fehlberg solver
(\texttt{RKF45}), using the parameters for yttrium iron garnet to define our
ferromagnetic layers.\footnote{We use the parameters from
  Ref.~\onlinecite{lan:2017aa}. Converting their parameters to our notation
  gives $\calJaf = 7.25\, \text{nm}^2\text{ps}^{-1}$, $2\calJf = 0.0221\,
  \text{ps}^{-1}$, $\lambda = 29.1\, \text{nm}$, $|\bD| = 0.663
  \text{nm}\,\text{ps}^{-1}$.} Our results are displayed and discussed in
Fig.~\ref{fig:single-wp-retarder}.  Note
  that, deep within the domain wall, the ``easy axis'' is no longer aligned with
  the textural slow mode, and the dispersion becomes imaginary for modes below a
  critical energy. In this case, spin transfered to the domain wall is the
  dominant process, and our numerical calculations break down close to this
  regime. Augmenting our theory with a collective coordinate theory of the
  domain wall, effectively allowing it to absorb spin, may be used to address
  this problem. Here, however, we keep the problem pedagogical by simply
  assuming that spin waves are sufficiently high energy that the local
  Hamiltonian remains Hermitian.

\label{sec:synthetic-afm-planar}
In our analysis of the domain wall retarder, we note an important
difference between the $g$-type and synthetic AFM in action.
Define $\mathcal C = \sigma_x\otimes\bbid_2$, which 
exchanges each underlying basis function $\{\alpha,\beta,\alpha^*,\beta^*\}$ with its conjugate (time-reversed) partner. This operation corresponds to \emph{charge
  conjugation}. $\mathcal C$ changes the sign of the coupling between
spin wave and the emergent electromagnetic fields arising from spin
texture and DMI. Together with time reversal (given by
complex conjugation), the full 
chirality operator $\mathcal S = \mathcal T\mathcal C$ is a symmetry of the degenerate
N\'eel-state Hamiltonian $\mathcal H = \hat h \oplus \hat h^*$. The breaking of
$\mathcal S$ symmetry by spin texture in the domain wall is what allows the
relative amplitudes of right- and left-handed modes to change in the overall wavefunction.

Now define $\mathcal I = \bbid_2\otimes\sigma_x$, which defines the sublattice
interchange operation. $\mathcal T\mathcal I$ is also a symmetry of the
degenerate Hamiltonian. In the $g$-type AFM case, spin texture will break $\mathcal
T\mathcal I$ symmetry in general, because an infintessimal misalignment is
present in each unit cell.\cite{Tveten:2015aa} In the SAF, however, the two
sublattice sites in a unit cell are never misaligned, so that $\mathcal
T\mathcal I$ is preserved even in the presense of spin texture.

Algebraically, the $\mathcal T\mathcal I$ symmetry of the SAF restricts
off-block-diagonal terms of the $4\times4$ spin wave Hamiltonian to be purely
real. Since the embedding $E$ is itself real, it follows that the isospin
Hamiltonian cannot have a nonzero $\tau_y$ component. The disentangling of
$\mathcal T\mathcal I$ from $\mathcal S$ symmetry in SAFs should be seen as a
virtue: it means that we can use SAFs to carry out rotations about precisely
known axes. By contrast, the $g$-type calculation in
Fig.~\ref{fig:single-wp-retarder} shows that symmetry-unconstrained rotations
can be quite complex. Not only is the axis of rotation not about a canonical
basis vector, but the axis of rotation \emph{changes dynamically} as the
wavepacket travels through the continuum of different local Hamiltonians
presented by the spin texture. Precise rotations appear to be insufferably
difficult to control in such an AFM, so our prescription to experimentalists and
device engineers is to use an SAF when precision is needed.
However, SAFs present their own challenges. Unlike pure
  $g$-type AFMs, SAFs present a shape anisotropy that may make the realization
  of uniaxial PMA difficult to maintain. A possible solution would be to use
  $a$-type AFMs. These materials are magnetically ordered at the lattice level,
  but are AFM-ordered in layers, rather than by nearest neighbors. These may
  present the best of both worlds: their symmetry constraints will disentangle
  different rotations, as with an SAF, but they would avoid shape anisotropy
  issues. Further materials research in this direction is warranted.

We emphasize that, although our wavepacket theory describes a single
semiclassical particle, it nonetheless applies to a global spin wave
state.\footnote{Though, naturally, it cannot express magnon-magnon
  interactions in that state.} Our results for both the domain wall
and the magnon FET match the micromagnetic simulations of
Refs.~\onlinecite{lan:2017aa} and \onlinecite{Cheng:2016kv} to within
five percent error in the driving frequency.\footnote{Since margins of
  error in these works are not given, we simply take their results as
  exact. Ref.~\onlinecite{lan:2017aa}, for instance, says that a 16.2
  GHz drive frequency produces a quarter wave plate, whereas our
  calculations require a 17.1 GHz drive---an error of $\sim 5\%$.}
Formally, the global wavefunction can be decomposed usefully into
wavepackets through a Gabor transformation. Standard signal analysis
indicates that this use of isospin wavepackets as a basis for the spin
wave signal is accurate as long as the grid spacing needed to sample
the spatially inhomogeneous texture does not exceed the spread of
wavelengths under consideration: $\Delta x_c\Delta k_c\leq 2\pi$.

\subsection{Other gates}

\begin{figure}
  \centering
  \includegraphics[width=\columnwidth]{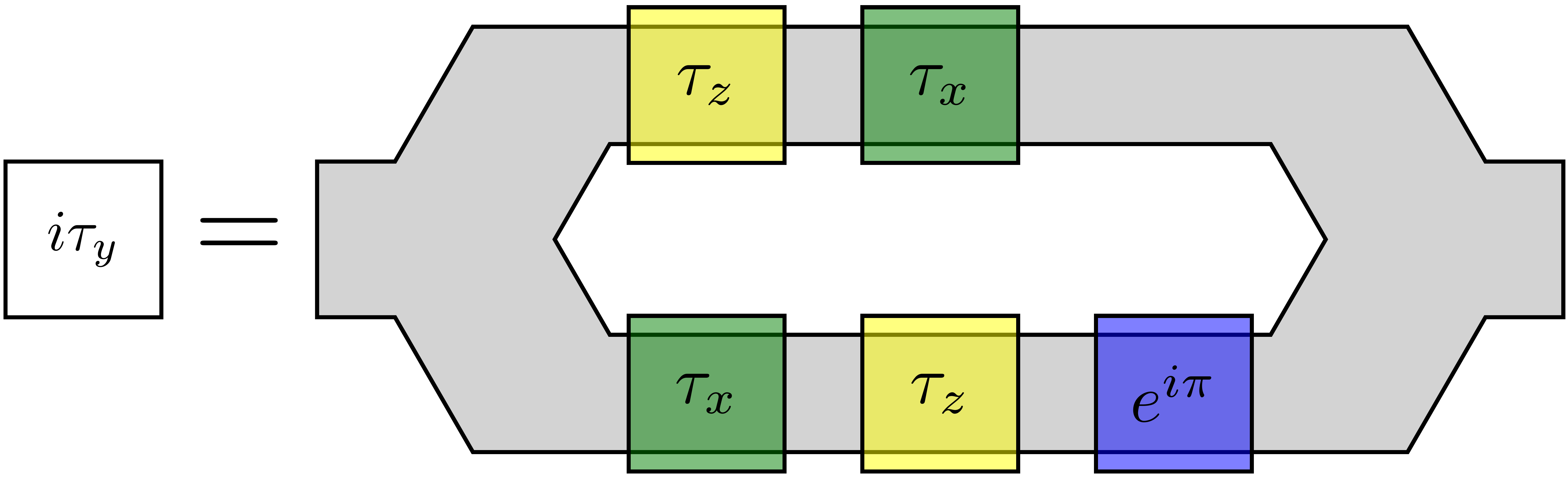}
  \caption{Applying different unitary gates to different branches of a
    spin wave signal makes the entire Lie algebra of rotational
    generators available from a set of two, as in the generation of
    $\sigma_y$ from the known $\sigma_x$ and $\sigma_z$ gates in the
    figure. By applying $\sigma_x$ on one branch and $\sigma_z$ on
    another, one could for instance generate a Hadamard
    gate. Note that in such a Hadamard gate, the designer must take care to
    ensure that the overall dynamical phase between the branches is equivalent,
    so as to avoid wave interference in the output channel. Since the U(1)
    phase is abelian, though, one need not worry about this in the $i\tau_y$
    gate pictured above. Using a $D_1 = -i\tau_z$ gate instead of a pure
    $\tau_z$ gate would generate the same, ``extraneous'' $\pi/2$ phase on both
    branches. \label{fig:nalc}}
\end{figure}

We have carried out explicit example calculations in the previous
sections because they can be immediately compared to results in the
literature, unifying these previous investigations under a single
formalism and allowing the reader to put our results in context.

However, our formalism is far-reaching and several other gates can be readily
designed. From straightforward calculations of $\mathcal H$ and $\mathscr H$,
one sees that a hard-axis anisotropy will provide a rotation about
$\sigma_x$.\footnote{In fact, this is not much of a surprise, since the SAF
  domain wall essentially acts as a local hard axis anisotropy, and we saw in
  Fig.~\ref{fig:single-wp-retarder} that the SAF domain wall acted as a
  $\sigma_x$ rotator as well.} Note that this actually implies spin
nonconservation, since the magnetization (relative to the local quantization
axis) carried by a spin wave corresponds to the polar angle of its isospin. Such
nonconservation mechanisms have been explored
elsewhere;\cite{khymyn:2016:transformation} here we merely accept that they fall
out of the isospin dynamical equations. Meanwhile, an applied magnetic field parallel with
the AFM order will provide a rotation about $\sigma_z$, since it breaks the
chiral degeneracy but not the U(1) symmetry of the ground state. In this way, a
parallel $\bB$ field gives the same effect as a normal $\bE$ field used to
generate the DMI in Sec.~\ref{sec:magfet}.

A local modification of the easy-axis anisotropy can raise or lower
the local AFMR frequency, and can therefore be used to adjust the
relative U(1) phase between two spin wave arms of a multi-channel
magnonic signal. For instance, such a modification could be used to
generate the blue $e^{i\pi}$ gate in Fig.~\ref{fig:nalc}. There, the
sign provided by the U(1) relative phase is crucial for computing the
commutator, rather than the anticommutator, of $\sigma_x$ and
$\sigma_z$---without the $e^{i\pi}$ gate, the loop in
Fig.~\ref{fig:nalc} would simply produce total destructive
interference, annihilating the input signal. If one could implement
this in a gate-controlled, switchable fashion, then electronic control
over the $e^{i\pi}$ gate (EAA) and the $\sigma_z$ gate (DMI) would
turn Fig.~\ref{fig:nalc} into a switchable
$\sigma_x \leftrightarrow \sigma_y$ gate. The presence of an
$e^{i\pi}$ gate allows multichannel schemes such as
Fig.~\ref{fig:nalc} to explore the full Lie algebra structure of
$\mathfrak{su}(2)$. Options for implementing a switchable $e^{i\pi}$
gate could include gate-controlled easy-axis anisotropy or a
perpendicular (to $\bn$) applied $\bB$-field. An especially important
use of this gate in a isospin computer would be to compensate the
accidental dynamical phase accumulated during the execution of
rotational gates.

If one is interested in investigating the effects of interactions not
considered here, one can simply derive the spin wave Hamiltonian in
the four-dimensional basis we have used in this paper and then project
it to the operator space over the degenerate subspace. One immediately
obtains the corresponding isospin Hamiltonian. We have tried to cover
the main classes of interactions in the Supplementary Material but
more unique interactions such as compass
anisotropy~\cite{Banerjee:2014eba} or honeycomb DMI~\cite{Cheng:2016ku}
could provide useful interfaces to other isospin operations.

\section{Discussion}

Our objective to this point has been to present the reader with a
cohesive program for isospin magnonics. We started by reviewing the
idea of chirality and the isospin vector that parameterizes it. Our
key foundational results were the semiclassical equations
Eqs.~\eqref{eq:sc-eoms} describing the isospin dynamics of an AFM
magnonic wavepacket. With these equation in hand, we described a
collection of physical gates---with a focus on voltage gates and
domain walls---that could manipulate the isospin in predictable,
calculatable ways. 

As this article draws to a close, let us reflect on our results and potential
avenues for future research. From the computing standpoint, recognition of the
chiral degree of freedom in AFM magnons is of paramount importance. Using the
isospin vector as a data carrier represents a paradigmatic improvement, on
multiple fronts, over the amplitude-modulating proposals that permeate FM
magnonics. First, power management and energy efficiency concerns that arise
when information is encoded in the FM spin wave power spectrum become immaterial
when the data is carried by AFM isospin. Many of the problems of architecture
scaling, which plague FM magnonic computing, are significantly alleviated in
AFMs. Second, the isospin carries a higher dimensionality of information. We
have seen that this considerably broadens the scope of magnon algorithmics.
For instance, it may be possible to replicate semiclassical
  quantum computing gates in isospin logic. If one is willing to accept the use
  of $2N$ isospin signals in place of $2^N$ qubits---and can map between these
  schemes faithfully---then perhaps one can ``classically simulate''
  non-entangling quantum circuits on a classical magnonic platform. To this end,
  a great deal of study is needed here to properly characterize the power and
scope of isospin computing.

Our key contribution to the field of magnonics is the development of a
generic, unified formalism for describing the isospin dynamics in
terms of unitary time evolution---a framework with which every
physicist is intimately familiar. Together with our mechanical recipe
Eq.~\eqref{eq:sw-ham-recipe} for generating the isospin Hamiltonian
from the free energy, we expect that our theory provides a cohesive
platform for future theoretical and experimental investigations into
the challenges of isospin magnonics.

Among these challenges are both extensions and applications of our
theoretical apparatus. The gates we investigated in
Sec.~\ref{sec:appl-select-magn} were purely one-dimensional, and from
these simple components one can produce quite sophisticated computing
devices. We have taken pains, however, to keep the spatial
dimensionality of our theory generic; one can apply the results of
this paper to 2D and 3D systems. Even in quasi-1D magnetic strips, two
dimensional textures such as skyrmions or magnetic vortices could
produce interesting effects. The interactions between such solitons
and AFM spin waves in open systems is also an open question. Our
theory could be used to address these issues. 

There of course exist magnonic applications outside the spin wave approximation
that underlies the theory in this paper. There, our technical theory may not be
a suitable tool, but we hope that our phenomenological description of the SU(2)
isospin---a concept which relies solely on a the fact that there are two
sublattice degrees of freedom with a relative phase between them---will prove
useful. Recently, for instance, AFM auto-oscillators have been
proposed.~\cite{cheng:2016aa, checinski:2017aa} The dynamical differences
between AFM and FM (Klein-Gordon versus Schr\"odinger) suggest that existing
theories of magnetic auto-oscillation~\cite{Slavin:2009ge} will
  need to be extended for the AFM case. This has already been done in the case
  of easy-plane oscillators, where the magnetization produced by an oscillation
  is relatively fixed.~\cite{khymyn:2017:antiferromagnetic} Other second order
  oscillator theories exist, but---especially once they become coupled---are
  often intractable.~\cite{acebron:2005:kuramoto, rodrigues:2016:kuramoto} They
  are also usually considered as phase oscillators. Whether these are the most
  natural theories for describing isospin oscillators is an open question.

In the AFM case, for instance, will the concept of an
auto-oscillation bandwidth extend to neighborhoods on the isospin
Bloch sphere? Such questions---which inherently depend on
nonlinearity---call for an understanding of isospin beyond the
harmonic spin wave regime. Along a different direction, the
adventurous theorist might consider extending our theory to an AFM of
more than two sublattices, attempting to derive the dynamics of an
SU($N$) isospin. 

Finally, we note that ferrimagnets satisfy conceptual prerequisites
for an SU(2) isospin, but are usually treated (in the YIG case, at
least) merely as low-damping FMs. Given the importance of ferrimagnets
to modern magnonics, a theoretical extension of our formalism to these
systems could be of immense interest. Though the two modes in
ferrimagnets would not be degenerate as they are in AFMs---and
therefore would require more energy for switching---one might still in
principle be able to carry out isospin logical operations. Research
into such systems could be critical for applied isospin computing.

We gratefully acknowledge Xiaochuan Wu, Yang Gao, Jin Lan, Vikesh Siddhu, and Zachary
McDargh for our insightful conversations. This work was supported by
the National Science Foundation (NSF), Office of Emerging Frontiers in
Research and Innovation under Award No.~EFRI-1433496 (M.W.D.), the NSF
East Asia and Pacific Summer Institute under Award No.~EAPSI-1515121
(M.W.D.), and the National Natural Science Foundation of China under
Grants No.~11722430 and No.~11474065 (W.Y.~and J.X.).

\appendix

\section{Temporal dynamics from the Berry phase Lagrangian}

\label{sec:quasistatic}
Though we introduced spin wave dynamics via Eqs.~\eqref{eq:llgs}, it
is possible to bypass the Landau-Lifshitz equation all
together. Instead we can appeal directly to the Lagrangian of our
classical field theory on $\alpha$ and $\beta$, given by
\begin{equation}
  L[\alpha,\beta,\bar\alpha,\bar\beta] = L^\text{BP} - F
\end{equation}
where $F$ is the magnetic free energy and $L^\text{BP}$ is the
so-called Berry phase Lagrangian.  The Berry phase Lagrangian is given
by
\begin{equation}
  L^\text{BP} = \frac{S}{\epsilon^d}\sum_{\Gamma}^{A,B}\int\frac{\bOmega_\Gamma \times \bmag_\Gamma}{1 - \bOmega_\Gamma}\cdot\frac{\dd \bmag_\Gamma}{\dd t} \dd[d]{x},
\end{equation}
where $\epsilon$ is the lattice constant and $\bOmega$ is the gauge
dependent orientation of the local Dirac string.~\cite{Altland:2010aa,
  frankel:2011aa, Tchernyshyov:2015fr} If one takes the variational
derivative of $L^\text{BP}$ by $\alpha$ and $\beta$, we will find the
left-hand side of Eq.~\eqref{eq:simple-4x4}. Even though we have
already arrived at this result from the perspective of the
Landau-Lifshitz equation, we repeat the derivation here using the
Lagrangian picture. We do so because the Lagrangian formalism should
be of greater generality and modularity,\footnote{For instance, one
  could add collective coordinate sectors to the Lagrangian we present
  here, and the new Lagrangian would then, in principle, encode
  magnon-soliton dynamics. We leave that procedure to future
  research.}  so that others may simply add terms to the Lagrangian
and repeat the process we are about to demonstrate.

Define $\lambda_A = \sqrt{1-2|\alpha|^2}$,
$\lambda_B = \sqrt{1-2|\beta|^2}$, and
$\lambda_m = \sqrt{1-2|\mu|^2}$. The basic idea in evaluating
$L^\text{BP}$ is simply to make the substitutions
\begin{subequations}
  \label{eq:central-sub}
\begin{align}
  R\ma &= \frac{\hat{x}}{\sqrt{2}}\left[ \alpha + \alpha^* + \lambda_A (\mu + \mu^*)\right]\nonumber\\
        &+ \frac{\hat{y}}{i\sqrt{2}}\left[ \alpha - \alpha^* + \lambda_A (\mu - \mu^*)\right]\nonumber\\
        &+ \hat{z}(\lambda_A\lambda_m - \alpha^*\mu - \mu^*\alpha)\label{eq:central-sub-a}\\
  R\mb &= \frac{\hat{x}}{\sqrt{2}}\left[ \beta + \beta^* + \lambda_B (\mu + \mu^*)\right]\nonumber\\
        &+ \frac{\hat{y}}{i\sqrt{2}}\left[ \beta - \beta^* + \lambda_B (\mu - \mu^*)\right]\nonumber\\
        &- \hat{z}(\lambda_B\lambda_m - \beta^*\mu - \mu^*\beta)\label{eq:central-sub-b}
\end{align}
\end{subequations}
into the Lagrangian and expand the result. The ``monolithic
substitutions'' Eqs.~\eqref{eq:central-sub} are derived in Appendix
\ref{app:mono}. So long as the Lagrangian is a linear operator on the
spin wave fields $\alpha$ and $\beta$, we end up with a
collection of terms
\begin{equation}
  L^\text{BP} = L^\text{BP}_0 + L^\text{BP}_1 + L^\text{BP}_2
\end{equation}
where we have collected terms at zeroth, linear, and quadratic order
in the spin wave fields. Linear spin wave theory, upon which our
formalism is built, cannot support terms at cubic order or higher, as
these would constitute nonlinearities in the equations of motion.

Because we are interested in taking functional derivatives with
respect to the spin wave fields, we can immediately neglect the terms
$L_0^\text{BP}$.\footnote{Such terms would need to be reserved for
  coupling to a Lagrangian collective coordinate theory of the
  textural dynamics, which we do not treat here.}  As for
$L_1^\text{BP}$, we see that functional derivatives of this term would
actually introduce inhomogeneous terms in the equations of motion. The
fastidious reader will find in her derivations that we apparently
\emph{do} have such terms in our Lagrangian, which do not vanish
\emph{a priori}. Such terms, if they properly belong to a physical
description of the system, would seem to imply spontaneous emission of
spin waves, since they will let $\dot\Psi$ take on a nonzero value
even when $\Psi$ is everywhere zero.

However, the reader is simultaneously invited to notice that we have
introduced more ``perturbations'' than we can actually control. The
problem is that $\covA{}$, which we treat as an independent field,
encodes the ground state of the system, as predetermined by anisotropy
and DMI. In fact, once the boundary conditions are given, $\covA{}$ is
strictly determined by these parameters.\footnote{To be precise,
  $\covA{}$ is strictly determined up to its topological charge,
  \emph{i.e.} up to the sectors defined in
  Ref.~\onlinecite{Belavin:1975uk}.} In equilibrium, one may compute
$\covA{}$ in principle by minimizing the free energy functional with
respect to the textural gauge fields,
\begin{equation}
  \label{eq:textural-constraint}
  \left\{\fdv{F[\bD,K ]}{A_\mu^j} = 0\right\}_{\mu,j}\implies \covA{\text{equilibrium}}[\bD,K ].
\end{equation}
Formally, these equations should be solved simultaneously with the
actual spin wave equation. On physical grounds, though, we assume that
these inhomogeneous terms always vanish when the system under
consideration is in equilibrium---or else, the system would not in
equilibrium, leading to a contradiction. The mathematical mechanism
transmitting this assumption is precisely the set of constraints
\eqref{eq:textural-constraint}. If the system is not in
equilibrium---say, if a soliton is moving---then generally speaking it
should generate spin waves inhomogeneously. Though our formalism
allows for temporal behavior of the underlying spin texture, we assume
that it is always in \emph{quasistatic equilibrium}---that is, we
neglect any inhomogeneous spin waves it generates.

After the above considerations are implemented, we find that we need
deal only with the harmonic Lagrangian
\begin{equation}
  \label{eq:quadratic-only-L}
  L^\text{BP} \mapsto L^\text{BP}_2.
\end{equation}
Keeping only the quadratic terms in the spin wave modes, keeping terms only to order $O(|\covA{}|^2)$ in our perturbative expansion, and summing over the sublattices $\Gamma \in \{A,B\}$, we are left merely with
\begin{align}
  L^\text{BP}_2 &= S\left[ A^z_t \bar\alpha \alpha + \frac{in}{2}\left(\bar\alpha\dot\alpha - \alpha\dot{\bar\alpha}\right)\right]\nonumber\\
  &-S\left[ A^z_t \bar\beta \beta + \frac{in}{2}\left(\bar\beta\dot\beta - \beta\dot{\bar\beta}\right)\right],
\end{align}
where $n = 1 + |\mu|^2$ is the effective index of refraction between
the local and vacuum values of the spin wave speed, as seen from the
Klein-Gordon formulation (see Appendix \ref{sec:staggered-order}). One
readily observes the difference of a minus sign separating sublattices
$A$ and $B$, as well as a minus sign between each field and its
conjugate partner. These signs are precisely our
$\tau_z\otimes\sigma_z$ factor from
Eq.~\eqref{eq:simple-4x4}. Defining
$\Psi = (\alpha,\beta,\bar\alpha,\bar\beta)$, we find that setting
$\delta L/\delta \bar\Psi = 0$ results in
\begin{equation}
  i (\tau_z\otimes\sigma_z) \dot\Psi = \frac{\epsilon^d}{n S}\frac{\delta F}{\delta \bar\Psi} - A_t^z(\bbid_2 \otimes\sigma_z)\Psi
\end{equation}
where $\epsilon$ is the lattice constant. Since we will only keep
the quadratic terms in $F$ by the arguments that lead to
Eq.~\eqref{eq:quadratic-only-L}, we know that $\delta F/\delta\Psi^*$
is a linear operation on $\Psi$ that can be written in the form
\begin{equation}
  i (\tau_z\otimes\sigma_z) \dot\Psi = \frac{\epsilon^d}{n S}\mathcal H \Psi- A_t^z(\bbid_2 \otimes\sigma_z)\Psi
  \label{eq:schro-4-general-app}
\end{equation}
analogous to Eq.~\eqref{eq:simple-4x4}. In general, the spin wave Hamiltonian is given by
\begin{equation}
  \label{eq:sw-ham-recipe}
  \mathcal H = \left(\begin{matrix}
      \ip{\fdv{F}{\alpha^*}}{\alpha} &  \ip{\fdv{F}{\alpha^*}}{\beta} & \ip{\fdv{F}{\alpha^*}}{\alpha^*} & \ip{\fdv{F}{\alpha^*}}{\beta^*} \\
      \ip{\fdv{F}{\beta^*}}{\alpha} &  \ip{\fdv{F}{\beta^*}}{\beta} & \ip{\fdv{F}{\beta^*}}{\alpha^*} & \ip{\fdv{F}{\beta^*}}{\beta^*} \\
      \ip{\fdv{F}{\alpha}}{\alpha} &  \ip{\fdv{F}{\alpha}}{\beta} & \ip{\fdv{F}{\alpha}}{\alpha^*} & \ip{\fdv{F}{\alpha}}{\beta^*} \\
      \ip{\fdv{F}{\beta}}{\alpha} &  \ip{\fdv{F}{\beta}}{\beta} & \ip{\fdv{F}{\beta}}{\alpha^*} & \ip{\fdv{F}{\beta}}{\beta^*} \\
      \end{matrix}\right),
\end{equation}
where the bra-ket notation simply indicates a functional inner product
under which the basis vectors $\alpha$, $\alpha^*$, $\beta$, and
$\beta^*$ are orthogonal.  Since the formula we have given for
$\mathcal H$ is explicit and straightforward---just make the
substitutions \eqref{eq:central-sub} into the free energy and start
taking functional derivatives of the quadratic sector---we will not
bore the reader with pages of algebra by deriving concrete
manifestations of $\mathcal H$ in the main text. We have provided
computer algebra code (in the Wolfram language) that derives
$\mathcal H$ for an assortment of useful free energies in the
Supplementary Material, with a focus on those free energies needed to
explore our various examples in Sec.~\ref{sec:appl-select-magn}. We
hope readers interested in their own systems will use the recipe
described above to generate their own spin wave Hamiltonians, which
project onto a Hamiltonian $\mathscr H$ governing the unitary dynamics
of the isospin vector $\ket{\bEta}$ in Sec.~\ref{sec:non-abel-wavep}.

\section{Spin texture}

\label{sec:spin-waves-spin}

A principal mechanism\footnote{Though, not the unique mechanism: see
  hard-axis anisotropy.} by which we break $U(1)$ symmetry and mix the
chiralities is through the introduction of a nonuniform ground
state. To that end, we require a formal structure for encoding
information about the ground state in our dynamical equations.

Much of the contemporary literature dealing with spin texture opts to
assemble a local coordinate frame, generally
$\{\hat e_r, \hat e_\theta, \hat e_\phi\}$, so that the linearization
process we used to derive Eq.~\eqref{eq:2-level-ham} can be recycled
in the $\{\hat e_\theta,\hat e_\phi\}$ plane. This amounts to a
passive transformation, taking the oscillatory plane of the spin wave
fluctuations to align with the local texture.

We instead opt to carry out the equivalent \emph{active
  transformation}, rotating each spin so that its spin wave plane
coincides with the global
$xy$-plane.\cite{Dugaev:2005aa,Guslienko:2010kq,Cheng:2012kl} A
thorough introduction to this technique in the ferromagnetic case is
given by Ref.~\onlinecite{Dugaev:2005aa}. In ferromagnets, one simply
defines a rotation matrix $\hat R(\bx,t)$ by
$\hat R \bmag^0 = \hat z$, so that it sends the ground state
configuration $\bmag(\bx,t)$ at each point to the global
$\hat z$-axis. This rotation matrix gives rise to a gauge field
$\covA{\mu} = (\partial_\mu \hat R)\hat R^T$. Formally, $\omega$ may
be regarded as a matrix-valued ($\mathfrak{so}(3)$-valued) one-form.

One can show in the lattice formalism that $\covA{}$ represents the
infinitesimal rotation $1 + R_iR_j^T = \exp \covA{ij} $ between two
sites, that is, $\covA{}$ is a generator of rotations. It can thus be
decomposed into the standard basis for $\mathfrak{so}(3)$,
\begin{equation}
  \covA{\mu} = A_\mu^x \hat J_x + A_\mu^y \hat J_y + A_\mu^z \hat J_z.
\end{equation}
Defining $R$ in terms of the Euler angles
$R = e^{-i\psi\hat J_z}e^{-i\theta\hat J_y}e^{-i\phi \hat J_z}$, we
can express the vector fields $\bA^j$ in terms of the spherical angles
describing the spin texture. This is why we have chosen to include
minus signs in the exponentials defining $R$: they show that we first
``undo'' the spherical angles by sending the azimuth to
$\phi - \phi = 0$ and then sending the polar angle to zero. Taking this convention gives us
\begin{subequations}
  \label{eq:explicit-A-vectors}
\begin{align}
  A^x_\mu &= -\sin\psi\,\partial_\mu\theta + \cos\phi \sin\theta\,\partial_\mu\phi,\\
  A^y_\mu &= -\cos\psi\,\partial_\mu\theta - \sin\theta\sin\psi\,\partial_\mu\phi,\\
  \text{and}\quad A^z_\mu &= -\cos\theta\,\partial_\mu\phi - \partial_\mu\psi.
\end{align}
\end{subequations}
Since only two angles are needed to specify the state of each spin,
the third rotation by $\psi$ appears to by extraneous, though
certainly permitted since it leaves invariant the spin texture now
lying along $\hat z$. In this sense, it represent the U(1) gauge
freedom associated with the U(1) symmetry of a coherent spin. In
practice, though, $\psi$ will often \emph{not} be a gauge freedom,
because the U(1) symmetry will often be broken by means other than the
immediate spin texture. If the spin texture has any misalignment with
the easy axis---that is, if there is \emph{any} deviation from the
Neel ground state---then the anisotropy energy will not be invariant
under the rotation by $\psi$. DMI or hard-axis anisotropy vectors
lying perpendicular to the ground state would also break this
symmetry.

The fact that we have chosen $\hat z$ as the global axis to which the
texture is rotated means that we will mostly be concerned with the
$J_z$ component of the curvature form $\bOmega = \dd \covA{}$. The
main consequence is that it is the curl of $\bA^z$, rather than the
curl of $\bA^x$ or $\bA^y$, which will provide the emergent
electromagnetic field---familiar to students of magnetic
skyrmions\cite{Nagaosa:2013aa}---generated by a spin texture.

The reader may recall that $\covA{}$, and therefore the 3-tuple
$(\bA^x,\bA^y,\bA^z)$, was supposed to describe an infinitesimal
rotation between neighboring spins. Such a rotation belongs to a two
dimensional group, and should be describable by exactly two numbers;
therefore we should seek a single constraint among our three vector
potentials $\bA^j$. By analyzing the curvature form, one can quickly
show that this constraint is
\begin{equation}
  \nabla\times\bA^z = \bA^x \times \bA^y.\label{eq:xyz-curl-times}
\end{equation}
Because $\hat z$ is privileged, it will be convenient to keep using
$\bA^z$ in our equations. For $\bA^x$ and $\bA^y$, though, we define a
more concise complex field via
\begin{equation}
  \calA_\mu = \frac{A^x_\mu + i A^y_\mu}{\sqrt 2}.
\end{equation}
Then we see that we can substitute the right-hand side of
Eq.~\ref{eq:xyz-curl-times} for\footnote{We are dealing only with the
  $\hat z$ component since we will generally not work in higher than
  $(2+1)$ dimension and therefore the magnetic field has only one
  component---hence the dot product by $\hat z$.}
\begin{equation}
  \hat z \cdot (\bA^x \times \bA^y) = A^x_x A^y_y - A^x_y A^y_x = 2i\calA^*_x\calA_y.
\end{equation}
We conclude that $\calA_x^*\calA_y$---and, therefore,
$\calA_y^*\calA_x$---is a physically interesting quantity, as it
encodes the same emergent electromagnetic field as the curl of $\bA^z$.

What of the symmetric products $\calA^*_\mu\calA_\mu$? It turns out
that these elements are also gauge invariant physical quantities. In
the general case, one finds
\begin{align}
  \calA_\mu^*\calA_\nu &= g_{\mu\nu} + \frac{i}{2} F_{\mu\nu},\\
  \text{defining}\quad Q_{\mu\nu} &= \calA_\mu\calA^*_\nu
\end{align}
where $g_{\mu\nu} = A^x_\mu A^x_\nu + A^y_\mu A^y_\nu$ reduces in
spherical angles of the texture to
\begin{align}
  g_{\mu\nu} &= \partial_\mu\theta\partial_\mu\theta + \sin^2\theta \partial_\mu\phi\partial_\nu\phi\\
  \implies g &= \dd{\theta}^2 + \sin^2\theta \dd{\phi}^2.
\end{align}
In other words, $g$ is just the first fundamental form on the
sphere. It is the differential line element $\dd s^2$ by which arc
lengths of the spin texture through spin space are measured. The
matrix $g$ is the spherical metric.

$Q_{\mu\nu}$ is called the \emph{quantum geometric
  tensor}.\index{quantum geometric tensor|textbf} There is very little
``quantum'' about it in our case, but the nomenclature is already out
there.\cite{cheng2010:qm, ma:2010aa, Piechon:2016ti, cheng:2017qm}

\section{A monolithic substitution for introducing the spin wave
  fields}
\label{app:mono}
In the antiferromagnetic case, we choose the rotation
matrix to send the staggered order to the global $\hat z$. Generally
speaking, $\ma$ and $\mb$ are not perfectly antiparallel, so after
this rotation we will still be left with in-plane components of the
(rotated) local magnetization.

We have already alluded to the fact that our two-level system does not
fully describe the spin wave dynamics. This is because the basis
fields $a_x+ia_y$ and $b_x+ib_y$ only represent circular modes. If we
want to access modes with linear components---say, fluctuations of
$a_x$ with $a_y = 0$, then our Hamiltonian needs to couple to a linear
combination of both $a_x + ia_y$ and its complex conjugate.

To address this, we have introduced the fields
$\alpha,\alpha^*,\beta,\;\text{and}\;\beta^*$ to represent our spin
wave fluctuations on each sublattice. We will abuse notation by
expressing the coefficients of each of these fields as
\begin{equation}
  \Psi = (\alpha,\beta,\bar\alpha,\bar\beta) \in \text{span}\{\alpha,\beta,\alpha^*,\beta^*\}
\end{equation}
Now let us fold these new variables into our formalism. First, split
each rotated field into its slow ($\tma^0$ and $\tmb^0$) and fast
($\balpha$ and $\bbeta$) modes---which are perpendicular by
construction---and then split the slow modes into the local staggered
order and local magnetization ($R\bn = \lambda_m \hat z$ and
$\tilde\bmag = R\bmag$, with $\lambda_m = \sqrt{1-m^2}$) of the
quasistatic equilibrium spin texture. We have introduced factors of
$\lambda_A = \sqrt{1-|\balpha|^2}$ and
$\lambda_B = \sqrt{1-|\bbeta|^2}$ in order to maintain the
normalization of the slow modes $\tma$ and $\tmb$ in the presence of
spin wave fluctuations. In other words, we have
\begin{subequations}
\begin{align}
  R\ma &= \lambda_A (\tilde\bmag + \lambda_m\hat z) + \balpha\\
  \text{and}\quad R\mb &= \lambda_B (\tilde\bmag - \lambda_m\hat z) + \bbeta
\end{align}
\end{subequations}
A few notes about the quantities we have just defined. First,
$\tilde\bmag$ lies in the $xy$-plane, since
$\tilde\bn = \lambda_m\hat z$ is perfectly out-of-plane. Second,
though we have opted out of a concern for brevity not to decorate
$\bn$ and $\bmag$ with any kind of indicator, keep in mind that these
variables \emph{only} encode the slow modes of the system. All spin
wave fluctuations of these quantities have been restricted by
construction to the excitations $\balpha$ and $\bbeta$.

Notice that we have chosen $R$ through the re-alignement of $\bn$ to
avoid choosing a preferred sublattice. Because each $\ma^0$ and
$\mb^0$ is subtle misaligned from $\bn$ in the presence of a texture,
however, our rotated spin wave fluctuations are $\balpha$ and $\bbeta$
have small out-of-plane components. It would be convenient instead to
restrict them to the $xy$-plane, so let us now compute exactly what
their out of plane component is. Since they are orthogonal to the
sublattice slow modes by construction, we have (on the $A$-sublattice,
for instance)
\begin{equation}
  0 = \balpha \cdot \tma^0= \balpha \cdot \tilde\bmag + \lambda_m\alpha_z
\end{equation}
so that $\alpha_z = -\lambda_m^{-1}\balpha\cdot\tilde\bmag$ and
$\beta_z = \lambda_m^{-1}\bbeta\cdot\tilde\bmag$. Defining $\ba$ and $\bb$
as the planar projections of the spin wave fields, we can then simply
write $\balpha = \ba - \lambda_m^{-1}(\ba\cdot\tilde\bmag)\hat z$ and so on.

Finally, we define complex variables $\alpha = (a_x + ia_y)/\sqrt 2$,
$\beta = (b_x + i b_y)/\sqrt 2$, and
$\mu = (\tilde m_x + i \tilde m_y)/\sqrt 2$. Taking all of these
definitions together, we have our two monolithic substitutions,
\begin{subequations}
  \label{eq:central-sub-app}
\begin{align}
  \tma &= \frac{\hat{x}}{\sqrt{2}}\left[ \alpha + \alpha^* + \lambda_A (\mu + \mu^*)\right]\nonumber\\
        &+ \frac{\hat{y}}{i\sqrt{2}}\left[ \alpha - \alpha^* + \lambda_A (\mu - \mu^*)\right]\nonumber\\
        &+ \hat{z}(\lambda_A\lambda_m - \alpha^*\mu - \mu^*\alpha)\label{eq:central-sub-a}\\
  \tmb &= \frac{\hat{x}}{\sqrt{2}}\left[ \beta + \beta^* + \lambda_B (\mu + \mu^*)\right]\nonumber\\
        &+ \frac{\hat{y}}{i\sqrt{2}}\left[ \beta - \beta^* + \lambda_B (\mu - \mu^*)\right]\nonumber\\
        &- \hat{z}(\lambda_B\lambda_m - \beta^*\mu - \mu^*\beta)\label{eq:central-sub-b}
\end{align}
\end{subequations}
With these quantities in hand, the free energy can be computed
explicitly, and by taking variations by $\alpha$ and $\beta$ of the
consequent Lagrangian, we can ultimately determine the spin wave
equation of motion.

A final note: we will generally take the field $\ma$ and $\mb$---and
therefore $\alpha$, $\alpha^*$, \emph{et cetera}---not as the
quantities we want to solve for, but the functional basis in which we
will expression solutions of the spin wave state. In other words,
solving for the wavefunction will operationally mean solving for the
coefficients weighting $\alpha$, $\alpha^*$, $\beta$, and
$\beta^*$---not for those basis functions themselves. To avoid
confusion, we will write the coefficients that we're actually solving
for as $\alpha$, $\bar\alpha$, $\beta$, and $\bar\beta$. In this way,
it is clear that $\alpha$ need not equal $\bar\alpha^*$, and in fact
generally will not. It is the basis vectors of these coefficients that
need to be conjugate, not the coefficients themselves.


\section{A more detailed discussion of non-abelian wavepacket theory}
\label{sec:nonab-wavep-theory}


Before computing a phase space Lagrangian governing the
semiclassical dynamics, we establish some self-consistency properties
of the wavepacket that will provide for useful identities during our
calculation.

\subsection{Normalization condition}

First, let us enforce a normalization condition on $|W\rangle$, given
by
\begin{equation}
  \langle W | \tau_z\otimes\sigma_z | W \rangle = 1.
\end{equation}
This leads to a normalization condition for the $\eta$, namely that
\begin{align}
  \langle W | \tau_z \sigma_z | W \rangle &= (-1)^j\int \dbq\dbk w_k^*w_q \eta_{j,k}^*\eta_{j,q}\langle \psi^j_k|\sigma_z|\psi^j_q\rangle\label{eq:eta-norm-1}\\
                                                    &= (-1)^{2j}\int\dbq |w_q|^2 |\eta_j|^2\label{eq:eta-norm-2}\\
  \implies  1 &= \langle \bEta | \bEta \rangle \label{eq:eta-normalization}
\end{align}
Eq.~\eqref{eq:eta-normalization} suggests that, unlike $\ket W$ and
$\ket\Psi$, $|\bEta\rangle$ will be subject to a traditional,
Euclidean \Schrodinger/ dynamics. Recall that the $\sigma_z$ inner
product in the two-level system did not provide a useful normalization
condition, as a result of the internal hyperbolic geometry. It is only
here in the four level system, where the signs from internal and
external geometries cancel each other, that we arrive at a
normalizable spin wave density (rather than spin density).

The calculational patterns from Eq.~\eqref{eq:eta-norm-1} sections the
internal derivations of wavepacket theory. We briefly outline the
logical flow of the computation for readers unfamiliar with the
formalism. The key stages needed to reduce any of our wavepacket inner
product are:
\begin{enumerate}
\item Use the fact that the wavevectors are ``block-diagonal'' (in the
  sense of Eq.~\eqref{eq:four-level-hyperbolic-sols}) to reduce the
  $\tau_z$ to a single $(-)^j$, and to avoid any cross terms between
  eigenvectors from different bands.
\item Establish an inner product of the internal band structure
  (\emph{e.g.}~$\mel*{\psi^j_k}{\sigma_z}{\psi^j_q}$). Extract the
  translation operators to find a factor of $\exp(i(\bq-\bk)\bx)$ and
  use the inner product---a real-space integral over the sample---to
  produce a $\delta^d(\bq-\bk)$.
\item Carry out one of the momentum space integrals to activate the
  Dirac delta function and reduce the problem to a single Brillouin
  zone. 
\item If the inner product from step 2 was a normalization condition
  of the internal geometry, then it produced a $(-)^j$ that, together
  with the sign from $\tau$, cancels to give positive
  unity. Otherwise, there is a nontrivial inner product
  $\mel{\bEta}{\hat O}{\bEta}$ that must be tracked.
\item Integrate by parts, use product rules, and use the normalization
  condition as necessary to manifest a factor of $|w_q|^2$ in the
  integrand. Interpret $|w_q|^2\mapsto\delta^d(\bq-\bqc)$ to carry out
  the final integral.
\end{enumerate}
Before evaluating the Lagrangian proper, we have one more useful
identity to compute: the expectation value of the position operator.

\subsection{Position operator}
\label{sec:position-operator}

Let us consider the self-consistency condition for the wavepacket
center. This means that we require the observable $\hat \bx$ to be
diagonal in the wavepacket basis, with eigenvalue $\bxc$ for
wavepacket $|W(\bxc,\bqc,\bEta,t)\rangle$. Therefore
\begin{equation}
  \langle W | (\tau_z\otimes\sigma_z)\hat{\bx}|W\rangle = \bxc\langle W|(\tau_z\otimes\sigma_z)|W\rangle
  \label{eq:xc-eigenvalue}
\end{equation}
The braket on the right then reduces to unity by the wavepacket
normalization.

Before we proceed, let us define the non-abelian Berry connection
\begin{equation}
  \fraka^j_\mu = \left( \begin{matrix}
      \ip{\Psi_q^0}{i\sigma_z\partial_\mu\Psi^0_q} & 0\\
      0 & -\ip{\Psi_q^1}{i\sigma_z\partial_\mu\Psi^1_q}
    \end{matrix}\right),
\end{equation}
wherein $\mu$ is a coordinate of the phase space dynamics. We will
therefore be concerned alternatively with $\fraka_\bx$, $\fraka_\bq$,
and $\fraka_t$. Calculating the left-hand side of
Eq.~\eqref{eq:xc-eigenvalue} using the matrix elements of the position
operator from Ref.~\onlinecite{blount1962formalisms}, we
find\footnote{An guide to carrying out type of calculation, albeit
  without the hyperbolic factors, can be found in the appendix of
  Ref.~\onlinecite{Sundaram:1999ht}.}
\begin{align}
  \bxc &= \langle W | (\tau_z\otimes\sigma_z) \hat \bx | W \rangle\\
       &= \langle\bm{\eta}|\fraka_{\bq}|\bm{\eta}\rangle+ \frac{\partial \gamma_c}{\partial \bq}.\label{eq:self-consistency}
\end{align}
In deriving Eq.~\eqref{eq:self-consistency}, we see our first example
of a non-cancellation between $\sigma_z$ and $\tau_z$. The Berry
connection is not \emph{merely} the normalization condition
$\mel{\Psi}{\sigma_z}{\Psi}$, and therefore cannot produce the sign
needed to cancel the $(-)^j$ factor. Instead, these signs have all
been contained within $\fraka$.

\subsection{Extracting the electromagnetic Lagrangian}
\label{sec:lorentz-force-from}

In Sec.~\ref{sec:spin-texture-maintext}, we introduced the collection of vector potentials
$\bA^j$ which encode the spin texture. Generally speaking, the
introduction of spin texture breaks the continuous translational
symmetry of the (continuum limit of the) \Neel/ ground state. Since
the $\bA^j$ are not necessarily gauge invariant, though, one expects
that the translational properties of the vector potentials need not
align in general with translational properties of the physical
system. The situation is similar to introducing an electromagnetic
vector potential in standard quantum mechanics; there, the
\emph{canonical} momentum operator $-i\partial_x$ must be adjusted to
the \emph{mechanical} momentum operator, $-i\partial_x -ieA$, where
only the latter is properly conserved.

Even without explicitly computing the spin wave Hamiltonian, we expect
that the kinetic energy term we explored in the two-level system will
appear to undergo a sort of Peierls substitution by $\bA^z$. With this
in mind, we will now perform a gauge transformation, removing the
$\bA^z$ from the kinetic energy terms and collecting it into a new
Lagrangian term which will completely encapsulate the emergent
electromagnetic interaction.

Define the matrix
\begin{equation}
  \label{eq:G-gauge-xform}
  \mathcal G = \exp\left[-i (\tau_z\otimes\bbid_2) (\bA^z \cdot\bx)\right].
\end{equation}
Then, inserting factors of $\mathcal G^\dagger \mathcal G$ into the Lagrangian, we have
\begin{equation}
  \mel**{W}{\calG^\dagger \calG \left(i\tau_z\sigma_z\frac{\dd}{\dd t} - \mathcal H - A^z_t\sigma_z\right)\calG^\dagger \calG}{W} \label{eq:gauged-wp-lagrangian}
\end{equation}
where the wavepackets and Hamiltonian are, at this point, still in the
original gauge choice, and the brackets represent the matrix element
of the operator on the diagonal in the wavepacket basis. To save
space, we have removed the explicit tensor product notation. We leave
it to the reader to interpret $\tau_z \mapsto \tau_z\otimes\bbid_2$
and $\sigma_z\mapsto\bbid_2\otimes\sigma_z$ as the context demands.

The value of the transformation by $\mathcal G$ is not only in an
internal simplification of $\mathcal H$ , but also in elegantly
extracting the emergent electromagnetic Lagrangian early in the
calculation. One readily sees after carrying out the time derivative
that the Lagrangian is
\begin{align}
  \mel**{\tilde W}{
      \left(-(\sigma_z\dot\bA^z\cdot\hat \bx)
      - A^z_t\sigma_z+ i\tau_z\sigma_z\frac{\dd}{\dd t}
      - \tilde{\mathcal H} \right)}{\tilde W} \label{eq:gauge-wp-lagrangian}
\end{align}
where $\tilde{\mathcal H} = \calG \mathcal H
\calG^\dagger$. Collecting the first two terms together, this can be
naturally split into three components:
\begin{equation}
  \label{eq:9}
  L = L_\text{EM} + L_{\dd t} + L_H.
\end{equation}
These components represent the emergent electromagnetic, dynamical,
and free energy sectors of the spin wave equation.

The gauge transformation has also affected the wavepacket
itself. Concretely, the wavepacket is now
\begin{align}
  \ket*{\tilde W} &:= \mathcal G\ket{W} = \int \dbq w(\bq,t)\big[\nonumber\\
    \tilde\eta_0(\bq,t) &\ket*{\Psi_0(\bq,t)}
    + \tilde\eta_1(\bq,t) \ket*{\Psi_1(\bq,t)}\big]
  \label{eq:gauged-wp}
\end{align}
where $\ket{\tilde\bEta} = (\tilde\eta_0,\tilde\eta_1)$ locates the
gauge-transformed wavepacket within the degenerate subspace.

From Eq.~\eqref{eq:gauge-wp-lagrangian}, we see the need to evaluate
\begin{equation}
  -\dot\bA^z\cdot\langle \tilde W|(\bbid_2\otimes\sigma_z)\hat\bx|\tilde W\rangle - A^z_t\mel*{\tilde W}{\bbid_2\otimes\sigma_z}{\tilde W}
\end{equation}
the first of which terms will invoke a calculation analogous to those
in Sec.~\ref{sec:position-operator}. We have
\begin{align}
  \langle \tilde W | (\bbid_2\otimes\sigma_z)\hat\bx|\tilde W\rangle
  &= \langle\bm{\eta}|\tau_z \fraka_{\bq}|\bm{\eta}\rangle+ \mel{\bEta}{\sigma_z}{\bEta}\frac{\partial \gamma_c}{\partial \bq}
\end{align}
Substituting in the self-consistency condition
Eq.~\eqref{eq:self-consistency} on $\bxc$ for the
$\gamma_c$-derivative, we end up with
\begin{align}
  \label{eq:em-lagrangian} 
  L_\text{EM} &= -\dot\bA^z\cdot\Gamma_{\bq} -\chi (\dot\bA^z\cdot\bxc + A^z_t)
\end{align}
where $\chi = \mel{\bEta}{\tau_z}{\bEta}$, and $\Gamma_{\bq}$ is the
covariance
$\mel{\bEta}{\tau_z\fraka_\bq}{\bEta} -
\mel{\bEta}{\tau_z}{\bEta}\mel{\bEta}{\fraka_\bq}{\bEta}$. Note that
we have simplified these terms back to $\bEta$, rather than
$\tilde\bEta$, since $\mathcal G$ commutes with $\tau_z$ and
$\fraka_\mu$.

Interpreting $\chi$ as a charge, the second half of
Eq.~\eqref{eq:em-lagrangian} is just the interaction Lagrangian for a
charged particle in an electromagnetic field.\cite{jackson:2007} Note
that the, in particle physics, there is also a sense in which the
electromagnetic charge is a $\tau_z$ expectation value: one can rotate
the isospin of a positively charged proton, through some SU(2)
``isospin'' space, to the neutrally charged neutron. That we have a
similar sort of continuum-valued (emergent) charge is our motivation
for employing the ``isospin'' nomenclature in our definition of
$\bEta$.
 
\subsection{Time derivative term}
Though we have already encountered a few time derivatives without
comment in the wavepacket theory, a few words are certainly in order
concerning the time variable. Its treatment is one of the most
delicate and subtle parts of wavepacket theory, and it is easy to make
dangerous systematic errors without a proper treatment. For the reader
interested in replicating our derivation, we have given some notes on
the matter in Appendix~\ref{sec:notes-time-deriv}.

The time derivative term $L_{\dd t}$ in the Lagrangian is
\begin{equation}
  i\int \,\dd\bq\,\dd\bk \,\langle \Psi^i_q|\tilde\eta^*_{i,q}w_q^*(\tau_z\sigma_z)\frac{\dd}{\dd t}\left(
    w_k\tilde\eta_{j,k}|\Psi^j_k\rangle
  \right)
    \label{eq:wp-time-deriv}
\end{equation}
Since our eigenvectors are themselves block diagonal, and since
$\tau_z\otimes\sigma_z$ as well as $G$ are both diagonal, we know
there can be no terms connecting $i\neq j$. 

The first term (on $w_k$) in a product rule of expansion of
Eq.~\eqref{eq:wp-time-deriv} is simply $\partial_t\gamma_c$.  The next
term, on $\tilde\eta_{j,k}$, generates the isospin dynamics, and the final
term gives rise to matrix-valued Berry connections. All together, these terms become
\begin{equation}
  L_{\dd t} = \mel{\tilde\bEta}{\dot\bx_c\cdot\fraka_\bx
    +  \dot\bq_c\cdot\fraka_\bq
    + \fraka_t 
    + i\partial_t}{\tilde\bEta} - \dot\bq_c\cdot\bxc
\end{equation}
We have used the self-consistency condition to replace the Berry phase
term $\partial_t\gamma_c$ with
$\bxc - \mel{\bEta}{\fraka_\bq}{\bEta}$.

\subsection{Hamiltonian terms}
\label{sec:hamiltonian-terms}
Finally, we have the terms coming from the spin wave Hamiltonian
itself. These are
\begin{align}
  \label{eq:sw-hamiltonian-wp-lagrangian}
  L_H &= -\langle W | \mathcal{H} | W \rangle \\
      &=- \frac{1}{ns}\int\dbq |w_q|^2\left[
        \tilde\eta_i^*\tilde\eta_j \mel{\Psi_i}{\tilde{\mathcal H}}{\Psi_j}
        \right]
\end{align}
Let us define the matrix
\begin{align}
  \tilde{\mathscr H} &= \left(\begin{matrix}
      \mel{\Psi^0_c}{\tilde{\mathcal H}}{\Psi^0_c} &
      \mel{\Psi^0_c}{\tilde{\mathcal H}}{\Psi^1_c} \\
      \mel{\Psi^1_c}{\tilde{\mathcal H}}{\Psi^0_c} &
      \mel{\Psi^1_c}{\tilde{\mathcal H}}{\Psi^1_c} 
      \end{matrix}\right).\label{eq:frak-h-def}
\end{align}
One may think of $\mathscr H$ as a projection of the original
$\mathcal H$ into the two-dimensional orthochronous degenerate
subspace that we are now calling ``isospin space''---the copy of SU(2)
in which $\bEta$ resides. Defining the embedding
\begin{equation}
  E^\dagger = \ket{\Psi_0} \bra 0 + \ket{\Psi_1} \bra{1} 
\end{equation}
which sends vectors in the isospin subspace to their representation in
parent $4\times 4$ Hilbert space space, $\mathscr H$ is merely
\begin{equation}
  \label{eq:ehe-dag}
  \tilde{\mathscr H} = E\tilde{\mathcal H} E^\dagger.
\end{equation}
This hermitian matrix will govern the dynamics of $\ket{\tilde\bEta}$
in the semiclassical dynamics we are about to describe.  The total
contribution from these energy terms to the wavepacket Lagrangian is,
simply,
\begin{equation}
  L_H = -\bra{\tilde\bEta} \tilde{\mathscr H} \ket{\tilde\bEta}.
\end{equation}

\subsection{Phase space EOMs}
Let's take stock of our progress. We a Lagrangian of three terms,
which have been reduced to
\begin{subequations}
  \label{eq:collected-wp-lagrangians-appendix}
  \begin{align}
    L_\text{EM} &=-\dot\bA^z\cdot\Gamma_{\bq}-\chi(\dot\bA^z\cdot \bxc + A^z_t),\label{eq:L-A-subeq}\\
    L_{\dd t} &= \mel{\tilde\bEta}{\dot\bx_c\cdot\fraka_\bx
    +  \dot\bq_c\cdot\fraka_\bq
    + \fraka_t 
    + i\partial_t}{\tilde\bEta} - \dot\bq_c\cdot\bxc\\
    L_H &=- \bra{\tilde\bEta} \tilde{\mathscr H} \ket{\tilde\bEta}.
  \end{align}
\end{subequations}
Now we can take variations against $\bxc$, $\bqc$, and $\ket\bEta$ to
derive semiclassical equations of motion.

First, let us find the force equation by taking a variation against
$\bxc$. For the Lorentz force term, we unsurprisingly have
\begin{align}
  \frac{\delta L_A}{\delta x^\mu_c}
  &= -\chi[\partial_t A^z_\mu
    +\partial_{x^\mu_c} A^z_t
    + \dot\bx_c\cdot\partial_{x^\mu_c}\bA^z
    - (\dot\bx_c\cdot\nabla)A^z_\mu]\\
  &= \chi \bE + \chi\dot\bx_c \times \bB
\end{align}
where we define the fields $\bE = -\nabla A^z_t -\partial_t\bA^z$ and
$\bB = \nabla \times \bA^z$ in the obvious ways. The time derivative
term meanwhile gives
\begin{align}
  \frac{\delta L_{\dd t}}{\delta x^\mu_c} 
  &= -\dot\bq_c + \langle\Omega_{\mu\nu}^{xx}\rangle\dot x^\nu_c + \langle\Omega^{xq}_{\mu\nu}\rangle\dot q^\nu_c + \langle\Omega_\mu^{xt}\rangle
\end{align}
where
\begin{equation}
  \langle\Omega^{\alpha\beta}_{\mu\nu}\rangle = \mel**{\bEta}{\left(\frac{\partial \fraka_{\beta^\nu}}{\partial \alpha^\mu} - \frac{\partial \fraka_{\alpha^\mu}}{\partial \beta^\nu}\right)}{\bEta}
\end{equation}
is the $\bEta{}$-density trace of the non-abelian Berry curvature, as
discussed in Ref.~\onlinecite{culcer:2005coherent}.

Finally, we have a contribution from the gauged Hamiltonian. In most
cases we consider in this paper, no such terms survive at
$O(|\covA{}|^2)$; the terms that might nominally survive are those
wrapped encoded in $L_\text{EM}$. Examples of terms that may survive
and not be included in $L_\text{EM}$ could include spatially dependent
anisotropy or DMI, arising from \emph{e.g.}~wedge-shaped layers in
magnetic heterostructures. Taking this term and the Lorentz force together, the force equation is
\begin{equation}
  \dot{\bq}_c = \text{Tr}\left[\hat\rho\left(\tau_z\left(\bE + \dot\bx_c\times\bB\right)-\frac{\partial \mathcal E}{\partial \bx_c} \right)\right]
  \label{eq:force-equation-wp}
\end{equation}
where $\mathcal E$ is the energy of the unperturbed degenerate bands,
and where we have defined the density operator
\begin{equation}
  \hat\rho = \rho_0\ket0 \bra0 + \rho_1 \ket1\bra1.
\end{equation}

Now we turn to the velocity equation. The results are little different
from what we would expect from standard non-abelian wavepacket theory,
giving us the classical velocity together with Berry-curvature induced
transverse velocities, 
\begin{align}
  \dot\bx_c =\text{Tr}\left[\hat\rho(\partial_\bq\mathcal E + \bOmega^{qq}\dot\bq_c + \bOmega^{qx}\dot\bx_c + \bOmega^{qt} )\right]
\end{align}
Now we turn to the most interesting equation of motion, generated by
the variation again $\bra{\bEta}$.  This generates terms of the form
\begin{align}
  \frac{\delta L_\text{EM}}{\delta \tilde{\bEta}^*}
  &= -\dot\bA^z\cdot \frac{\delta \Gamma_{\bq}}{\delta \tilde{\bEta}^*}-(\dot\bA^z\cdot\bxc)\tau_z\tilde{\bEta} - A^z_t\tau_z\tilde{\bEta} \\
  \frac{\delta \Gamma_{\bq}}{\delta \tilde{\bEta}^*} &= \tau_z\fraka_\bq \tilde{\bEta} - \tau_z(\mel{\tilde{\bEta}}{\fraka_\bq}{\tilde{\bEta}})\tilde{\bEta} - \chi\fraka_\bq\tilde{\bEta}\\
  \frac{\delta L_{\dd t}}{\delta\tilde{\bEta}^*}
  &= \left[
    \dot\bq_c\cdot\fraka_\bq
    + i\frac{\partial}{\partial t}
    \right] \tilde{\bEta}\\
  \frac{\delta L_H}{\delta\tilde{\bEta}^*} &= -\mathscr H \tilde{\bEta}
\end{align}
The final general equation of motion is
\begin{equation}
 i\left(\frac{\dd}{\dd t} + \mathscr{A}_t\right)\bEta
  = \left[\mathscr{H}+ \tau_z A^z_t + \hat{V}_\chi\right]\bEta
\end{equation}
where $\mathscr H = E\mathcal H E^\dagger$ (note that we have removed
the gauge transformation $\mathcal G$), $\mathscr A$ is the time
covariant connection on phase space,
\begin{align}
  \mathscr{A}_t &= \dot\bq_q\cdot\fraka_\bq + \dot\bx_c\cdot\fraka_\bx + \fraka_t \\
  \mathscr{A}_t^{ij} &= \mel**{\psi^i_c}{i\sigma_z\frac{\dd}{\dd t}}{\psi^j_c},
\end{align}
and $\hat{V}_\chi$ is a nonlinear term deriving from
$\Gamma_\bq$. it is given by
\begin{equation}
  \hat{V}_\chi = -\dot{\bA}^z\cdot\left(\tau_z\fraka_\bq - \tau_z \hat P_\eta \fraka_\bq - \fraka_\bq \hat P_\eta \tau_z\right),
\end{equation}
where $\hat P_\eta = \ket{\bEta}\bra{\bEta}$ is the projector onto the
isospin state, and as before the dot product (with $\dot\bA^z$) is
taken with the subscript in $\fraka_\bq$. This potential is nonlinear
in the sense that, through $\hat P_\eta$, it depends quadratically on
the current state, and the resulting term in the Hamiltonian has the
schematic form $|\psi|^2\ket\psi$. However, this nonlinear term
balances precipitously on the edge of irrelevance. $\dot\bA^z$ is
itself $O(\covA{}^2)$, so this term survives only if $\fraka_\bq$ is
$O(\covA{}^0)$. Though there is no reason (to our knowledge) this
could not happen in principle, none of the concrete systems we
consider later in the paper can activate this term. What's more, the
term would seem only relevant in the case of a moving spin texture, so
that $\dot{\bA}^z$ is nonzero. Such a term may be of interest for
those working in the dynamics of AFM solitons, but we leave that to
future research.
\subsection{Notes on time derivatives in wavepacket theory}
\label{sec:notes-time-deriv}
There are two variables, $\bx$ and $\bq$, that have been floating
around as dummy variables of integration in some of our
calculations. Several functions, such as the wavepacket envelope
$a_q=a(\bq,t)$ or the Bloch eigenvectors $e^{i\bq\bx}u(\bq,t)$, are
functions of both $\bq$ (or $\bx$) and time. For these functions,
there is no difference between a total time derivative and a partial
time derivative, because $\bq$ and $\bx$ are clearly independent
variables---merely coordinates of a space---that do not, themselves,
possess any temporal dynamics.

On the other hand, the gauge field $\bA^z=\bA^z(\bxc,t)$ was
originally, and will always be, evaluated at $\bxc$ in its spatial
argument. This $\bxc$ is a dynamical variable, which does depend on
time and has dynamics. Our notation, which follows
Ref.~\onlinecite{Sundaram:1999ht}, is that a partial time derivative
of such a function acts only on the second argument slot, where there
is an explicit time dependence. A total time derivative, on the other
hand, would include the time dependence through $\bxc$, so that
\begin{equation}
  \frac{\dd{}}{\dd t} = \pdv{t} + \dot\bx_c \cdot \pdv{\bxc}.
\end{equation}
So far our discussion has perhaps clarified the notation, but is by no
means unusual. The delicacy of these operations in wavepacket theory
occurs when evaluation of a wavepacket expectation value promotes a
function in the integrand---where it may have possessed only an
explicit time dependence---to a function of $\bqc(t)$, due to the
firing of the Dirac delta function $|a_q|^2$. The question is: should
the time derivative under the integrand be lifted to a total time
derivative or a partial time derivative once the function acquires a
new time dependence in the phase space coordinate arguments $\bxc(t)$
and $\bqc(t)$?

The answer is that we must promote it to a partial time
derivative. The original, physical meaning of such a time derivative
in the integrand was to ask how, at any given point in space, a
function changed with time. We are concerned with \emph{the function's
  temporal behavior}, not the temporal behavior of the combined
wavepacket-function system. From a different perspective, we note that
we are certainly free to take the time-derivative as early as
possible. Suppose we ``carry out'' the time derivative in the
integrand by replacing $\partial_t f(t,\bq)$ with its formal
derivative $F(t,\bq)$. Now $F$ is just a function which we have
determined in principle before ever introducing the phase space path
$(\bxc,\bqc)$, so after firing the Delta function we simply have
$F(t,\bqc(t))$. Clearly $F(t,\bqc(t)) = \partial_t f(t,\bqc(t))$, with
the derivative only in the first argument.

\section{Staggered order}
\label{sec:staggered-order}
Suppose we changed the basis of Eq.~\eqref{eq:2-level-ham} by a Hadamard matrix
\begin{equation}
  M = \frac{1}{2}\left(\begin{matrix}
      1 & 1 \\
      1 & -1 \end{matrix}\right),
\end{equation}
sending $\alpha$ and $\beta$ to $\delta m = m_x + i m_y$ and
$\delta n = n_x + i n_y$, respectively. Neglecting anisotropy for the
moment, the resulting Schrodinger equation on $\hat h$ is
\begin{equation}
  i\sigma_x \dv{t}
  \left[\begin{matrix} \delta m \\ \delta n \end{matrix}\right]
  = \frac{1}{2}\left(\calZ + \sigma_z(\calZ - \calJ\nabla^2)\right)
  \left[\begin{matrix} \delta m \\ \delta n \end{matrix}\right].
\end{equation}
Neglecting the dynamics of $\delta m$, this can be solved by taking a
second time derivative and the plugging the original equation for
$\dot n$ into the new equation for $\ddot n$. The result is
\begin{equation}
  \label{eq:classical-wave}
  0 = \left(\frac{1}{c^2}\dv[2]{t} - \nabla^2\right)\delta n
\end{equation}
where $c = \sqrt{\calZ\calJ/2}$. Therefore our $\sigma_z$-measured
\Schrodinger/ dynamics given (in either of the equivalent $2\times 2$ blocks) by Eq.~\eqref{eq:simple-4x4} are in
fact equivalent---in this simple regime, at least---to the
Klein-Gordon-type second order dynamics found more commonly in the
literature. It is no surprise that relativistic dynamics describe the
system whose modes, as we have seen, are restricted to timelike points
in a hyperboloid of two sheets. In the case where $K  = 0$, we
actually have massless particles, the hyperboloid of two sheets
becomes a light cone. Adding the anisotropy restores a mass term
$(\square - m^2)\delta n = 0$ to the KG equation, just as it opens a
mass gap in our hyperboloid.


%
\end{document}